\pdfoutput=1
\documentclass[11pt,a4paper]{article}

\usepackage{amsmath,amssymb,authblk,multirow,collref}
\usepackage{graphicx}
\usepackage{xparse}
\usepackage{subcaption}
\usepackage{color}
\usepackage{geometry}
\usepackage{booktabs}

\newcommand{\braket}[1]{\ensuremath{\left\langle#1\right\rangle}}

\usepackage{pgfplots}
\usepackage{xparse}

\usepackage{etoolbox}
\makeatletter
\patchcmd{\@sect}{#8}{\boldmath #8}{}{}
\let\ori@chapter\@chapter
\def\@chapter[#1]#2{\ori@chapter[\boldmath#1]{\boldmath#2}}
\makeatother

\usepackage[sort&compress,numbers,colon]{natbib}
\usepackage{hyperref}
\hypersetup{
pdfauthor={Yi Cai and Michael A. Schmidt},
pdftitle={The Minimal RnuMDM Model}
}

\usetikzlibrary{positioning,arrows,patterns}
\usetikzlibrary{decorations.pathmorphing}
\usetikzlibrary{decorations.markings}
\usetikzlibrary{calc}
\usetikzlibrary{shapes.misc}
\tikzset{
    photon/.style={decorate, decoration={snake}, draw=black, thick},
    fermionnoarrow/.style={draw=black, postaction={decorate}, thick},
    scalar/.style={draw=black, postaction={decorate}, decoration={markings,mark=at position .55 with {\arrow{>}}}, thick, dashed},
    scalarnoarrow/.style={draw=black, postaction={decorate},  thick, dashed},
    fermion/.style={draw=black, postaction={decorate},decoration={markings,mark=at position .55 with {\arrow{>}}}, thick},
    gluon/.style={decorate, draw=black, decoration={coil,amplitude=4pt, segment length=5pt}, thick},
    vertex/.style={draw,shape=circle,fill=black,minimum size=3pt,inner sep=0pt},
    cross/.style={cross out, draw=black,thick, minimum size=6pt, inner sep=0pt, outer sep=0pt}
}
\NewDocumentCommand\semiloop{O{black}mmmO{}O{above}}
{%
\draw[#1] let \p1 = ($(#3)-(#2)$) in (#3) arc (#4:({#4+180}):({0.5*veclen(\x1,\y1)})node[midway, #6] {#5};)
}
\NewDocumentCommand\lowsemiloop{O{black}mmmO{}O{above}}
{%
\draw[#1] let \p1 = ($(#3)-(#2)$) in (#2) arc (#4:({#4-180}):({0.5*veclen(\x1,\y1)})node[midway, #6] {#5};)
}
\NewDocumentCommand\upsemiloop{O{black}mmmO{}O{below}}
{%
\draw[#1] let \p1 = ($(#3)-(#2)$) in (#2) arc (#4:({#4+180}):({0.5*veclen(\x1,\y1)})node[midway, #6] {#5};)
}

\begin{document}
\title{Revisiting the R$\nu$MDM Models}
\author[1]{Yi Cai\thanks{\texttt{yi.cai@unimelb.edu.au}}}
\author[2]{Michael A. Schmidt\thanks{\texttt{michael.schmidt@sydney.edu.au}}}
\affil[1]{ARC Centre of Excellence for Particle Physics at the Terascale, School of Physics,
The University of Melbourne, Victoria 3010, Australia}
\affil[2]{ARC Centre of Excellence for Particle Physics at the Terascale, School of Physics, 
The University of Sydney, NSW 2006, Australia}
\date{}
\maketitle

\begin{abstract}
Combining neutrino mass generation and a dark matter candidate in a unified model has always been intriguing.   
We revisit the class of R$\nu$MDM models, which incorporate minimal dark matter in
radiative neutrino mass models based on the one-loop ultraviolet completions
of the Weinberg operator. The possibility of an exact accidental $Z_2$ is
completely ruled out in this scenario. We study the phenomenology of one of the models
with an approximate $Z_2$ symmetry.
In addition to the Standard Model particles, it contains two real scalar quintuplets, one vector-like quadruplet fermion and a
fermionic quintuplet.
The neutral component of the fermionic quintuplet serves as a good dark matter candidate which can be tested 
by the future direct and indirect detection experiments. 
The constraints from flavor physics and electroweak-scale naturalness are also discussed.  
\end{abstract}

\section{Introduction}
The particle identity of dark matter is one of the most important problems  
in physics beyond the Standard Model (SM). Generally in dark matter models a discrete symmetry, 
$Z_2$ symmetry as the simplest example,
has to be imposed by hand or show up as the remnant symmetry of a larger group
to protect the dark matter particle from decaying.    
Alternatively, the SM gauge group can be used to stabilize dark matter as discussed in minimal dark matter 
(MDM)~\cite{Cirelli:2005uq,Cirelli:2009uv} models. Higher representations of SU(2)$_L$ are introduced in MDM models as dark matter candidates. 
They couple to the SM sector only through gauge interactions, while other types of interactions such as 
Yukawa interactions or scalar interactions are all forbidden by SU(2)$_L$ due to
their large dimensions\footnote{Higher dimensional operators, which might be
present, can still lead to dark matter decay. See Ref.~\cite{DelNobile:2015bqo}
for a recent discussion. }.
As a result, an {\it accidental} (approximate) $Z_2$ symmetry is present. The
symmetry might be exact or approximately realised if the lifetime is larger than
the age of the Universe.

If we consider more than just a single SU(2)$_L$ multiplet, we might be able to
write down an interaction between SM fermions or Higgs and a pair of dark sector particles.    
It is of great interests to see whether it is possible to simultaneously 
explain the origin of neutrino masses and mixings, another clear evidence for physics beyond the SM. 
There is a rich literature on dark matter in neutrino mass
models (See
Refs.~\cite{Dodelson:1993je,Krauss:2002px,Ma:2006km,Basso:2012ti,Law:2013saa,Ahriche:2014cda} for
example.). The usual seesaw mechanisms~\cite{Minkowski:1977sc, Yanagida:1980,
Glashow:1979vf, Gell-Mann:1980vs, Mohapatra:1980ia, Magg:1980ut,
Schechter:1980gr, Wetterich:1981bx, Lazarides:1980nt, Mohapatra:1980yp,
Cheng:1980qt, Foot:1988aq} obviously can not be incorporated in the framework of MDM, 
as the intermediate particles couple to a pair of SM particles and thus are even under the accidental $Z_2$ of MDM.
Therefore we turn our eyes to the next-to-minimal solution, radiative neutrino
mass generation~\cite{Zee:1980ai}. We will focus on minimal ultraviolet (UV)
completions of the Weinberg operator.    

This idea of realizing radiative neutrino mass in a minimal dark matter model, coined as R$\nu$MDM, was proposed in
Ref.~\cite{Cai:2011qr},
where a scalar sextet and a fermionic quintuplet are introduced. The neutrino mass would be generated at one-loop level.
However, the accidental $Z_2$ symmetry is broken by a quartic coupling of three
scalar multiplets with one Higgs~\cite{Kumericki:2012bf} and it can be only
approximately realised in the limit of a small quartic coupling.
There are also attempts at higher loop order (See e.g.
Ref.~\cite{Ahriche:2014oda,Culjak:2015qja,Ahriche:2015wha}) and based on the generalized Weinberg operator~\cite{Kumericki:2012bh}.
Now we would like to take a second look at the one-loop completions of the Weinberg operator and
perform a systematic study and explore the possibility of R$\nu$MDM.     

This work is organized as follows:  In Sec.~\ref{sec:top} we discuss the
one-loop topologies and determine the possible R$\nu$MDM model realizations. In
Sec.~\ref{sec:model} we discuss one R$\nu$MDM model and obtain an
expression for neutrino mass.
The relevant constraints from Higgs and flavor physics are elaborated on in
Sec.~\ref{sec:flavor} and electroweak-scale naturalness in
Sec.~\ref{sec:naturalness}.
In Sec.~\ref{sec:other} we discuss the allowed parameter space as well as other
phenomenological issues, like collider phenomenology. We conclude in
Sec.~\ref{sec:con}. Technical details are collected in the appendices.

\section{One-Loop R$\nu$MDM Models}
\label{sec:top}
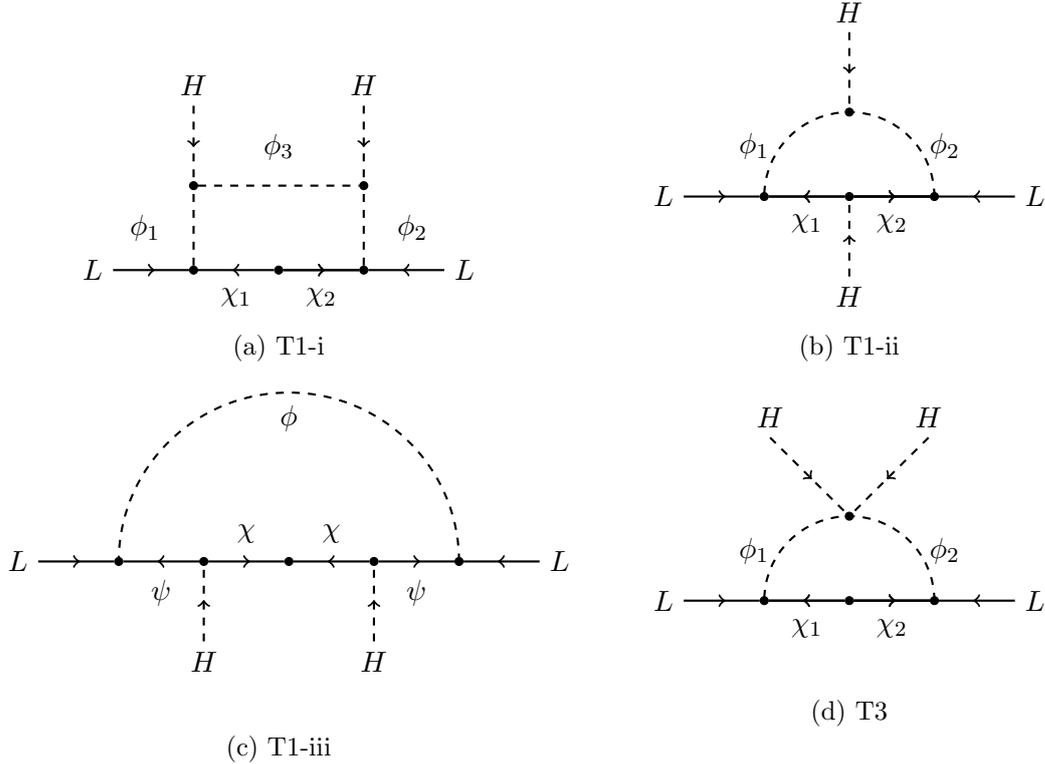
\begin{figure}[t!]
\centering
    \begin{subfigure}[t] {0.45\textwidth}
        \centering
        \begin{tikzpicture}[node distance=1cm and 1cm]
        \coordinate[ label=left:$L$] (v1);
        \coordinate[vertex, right=of v1] (v2);
        \coordinate[vertex, right=of v2] (v3);
        \coordinate[vertex, above=of v2] (v6);
        \coordinate[vertex, right=of v3] (v4);
        \coordinate[vertex, above=of v4] (v7);
        \coordinate[ right=of v4, label=right:$L$] (v5);
        \coordinate[ above=of v6, label=above:$H$] (v8);
        \coordinate[ above=of v7, label=above:$H$] (v9);    
        \draw[fermion] (v1) -- (v2) ;
        \draw[fermion] (v5) -- (v4) ;
        \draw[scalar] (v9) -- (v7) ;
        \draw[scalar] (v8) -- (v6);
        \draw[fermion] (v3) -- (v4) ;
        \draw[fermion] (v3) -- (v2) node[midway, below=0.1cm] {$\chi_1$};
        \draw[scalarnoarrow] (v2) -- (v6) node[midway, left=0.3cm] {$\phi_1$};
        \draw[fermion] (v3) -- (v4) node[midway, below=0.1cm] {$\chi_2$};
        \draw[scalarnoarrow] (v4) -- (v7) node[midway, right=0.3cm] {$\phi_2$};
        \draw[scalarnoarrow] (v6) -- (v7) node[midway, above=0.2cm] {$\phi_3$};       
        \end{tikzpicture}
	\caption{T1-i}
        \label{fig:nmassc}
     \end{subfigure}
\begin{subfigure}[t]{0.45\textwidth}
        \centering
        \begin{tikzpicture}[node distance=1cm and 1cm]
        \coordinate[ label=left:$L$] (v1);
        \coordinate[vertex, right=of v1] (v2);
        \coordinate[vertex, right=of v2] (v3);
        \coordinate[vertex, above=of v3] (v6);
        \coordinate[above=of v6, label=above:$H$] (v7);
        \coordinate[below=of v3, label=below:$H$] (v8);
        \coordinate[vertex, right=of v3] (v4);
        \coordinate[ right=of v4, label=right:$L$] (v5);
        \draw[fermion] (v1) -- (v2) ;
        \draw[fermion] (v5) -- (v4) ;
        \draw[scalar] (v7) -- (v6) ;
        \draw[scalar] (v8) -- (v3) ;
        \semiloop[scalarnoarrow]{v2}{v4}{0} ;
        \draw[fermion] (v3) -- (v2) node[midway, below=0.1cm] {$\chi_1$};
        \draw[fermion] (v3) -- (v2) node[midway, above left=0.5cm] {$\phi_1$};
        \draw[fermion] (v3) -- (v4) node[midway, below=0.1cm] {$\chi_2$};
        \draw[fermion] (v3) -- (v4) node[midway, above right=0.5cm] {$\phi_2$};
        \draw[fermion] (v3) -- (v4) ;
        \end{tikzpicture}
	\caption{T1-ii}
        \label{fig:nmassa}
     \end{subfigure}

     \vspace{2ex}

    \begin{subfigure}[t]{0.45\textwidth}
        \centering
        \begin{tikzpicture}[node distance=1cm and 1cm]
        \coordinate[ label=left:$L$] (v1);
        \coordinate[vertex, right=of v1] (v2);
        \coordinate[vertex, right=of v2] (v3);
        \coordinate[vertex, right=of v3] (v4);
        \coordinate[vertex, right=of v4] (v5);
        \coordinate[vertex, right=of v5] (v6);
        \coordinate[ right=of v6, label=right:$L$] (v7);
        \coordinate[below=of v3, label=below:$H$] (v8);
        \coordinate[below=of v5, label=below:$H$] (v9);
        \draw[fermion] (v1) -- (v2) ;
        \draw[fermion] (v7) -- (v6) ;
        \draw[scalar] (v8) -- (v3) ;
        \draw[scalar] (v9) -- (v5) ;
        \semiloop[scalarnoarrow]{v2}{v6}{0};
	\node at ($(v4)+(0,1.9)$) {$\phi$};
        \draw[fermion] (v3) -- (v2) node[midway, below=0.1cm] {$\psi$}; 
        \draw[fermion] (v3) -- (v4) node[midway, above=0.1cm] {$\chi$}; 
        \draw[fermion] (v5) -- (v4) node[midway, above=0.1cm] {$\chi$}; 
        \draw[fermion] (v5) -- (v6) node[midway, below=0.1cm] {$\psi$}; 
        \end{tikzpicture}
        \label{fig:nmassd}
	\caption{T1-iii}
    \end{subfigure}
    \begin{subfigure}[t]{0.45\textwidth}
        \centering
        \begin{tikzpicture}[node distance=1cm and 1cm]
        \coordinate[ label=left:$L$] (v1);
        \coordinate[vertex, right=of v1] (v2);
        \coordinate[vertex, right=of v2] (v3);
        \coordinate[vertex, above=of v3] (v6);
        \coordinate[above left=of v6, label=above:$H$] (v7);
        \coordinate[above right=of v6, label=above:$H$] (v9);
        \coordinate[below=of v3] (v8);
        \coordinate[below=of v8] (v10);
        \coordinate[vertex, right=of v3] (v4);
        \coordinate[ right=of v4, label=right:$L$] (v5);
        \draw[fermion] (v1) -- (v2) ;
        \draw[fermion] (v5) -- (v4) ;
        \draw[scalar] (v7) -- (v6) ;
        \draw[scalar] (v9) -- (v6) ;
        \draw[white] (v3) -- (v8) ;
        \semiloop[scalarnoarrow]{v2}{v4}{0};
        \draw[fermion] (v3) -- (v4) ;
        \draw[fermion] (v3) -- (v2) node[midway, below=0.1cm] {$\chi_1$};
        \draw[fermion] (v3) -- (v2) node[midway, above left=0.5cm] {$\phi_1$};
        \draw[fermion] (v3) -- (v4) node[midway, below=0.1cm] {$\chi_2$};
        \draw[fermion] (v3) -- (v4) node[midway, above right=0.5cm] {$\phi_2$};
        \end{tikzpicture}
	\caption{T3}
        \label{fig:nmassb}
     \end{subfigure}

\caption{Irreducible one-loop topologies for minimal UV completions of the
Weinberg operator.}
\label{fig:nmass}
\end{figure}

Majorana neutrino masses can be conveniently expressed in the form of the Weinberg operator 
\begin{equation}
\mathcal{O}_{\rm Weinberg} = \frac{LLHH}{\Lambda},
\end{equation}
where $\Lambda$ is the suppression scale and $L$ and $H$ denote the lepton and the Higgs doublets in the SM.
The minimal UV completions of the Weinberg operator at tree level are the seesaw mechanisms.
However the Weinberg operator can also be UV completed at loop level and
neutrino mass is generated radiatively. 
A systematic study of the one-loop UV completions has been performed in Ref.~\cite{Bonnet:2012kz}.
Part of the one-loop completions also induce one of the seesaw mechanisms whose contribution is mostly dominant.
These models are thus reducible and not genuine one-loop models.  
We show all the topologies that allow irreducible one-loop completions in Fig.~\ref{fig:nmass}. 
The well-known radiative seesaw model~\cite{Ma:2006km} is a realization of T3
shown in Fig.~\ref{fig:nmassb}. 
We want to first check the possibility to accommodate MDM in these topologies.
A good R$\nu$MDM model should at least satisfy the following three conditions:
\begin{enumerate}
\item At least one of the exotic particles in the loop should be a plausible MDM candidate,
which can be either a fermionic quintuplet or a scalar septuplet with zero hypercharge~\cite{Cirelli:2009uv}.
\item Any exotic scalar with hypercharge $\pm \frac{1}{2}$ has to be in an
	odd-dimensional SU(2) representation. Otherwise the quartic term
	$\phi^\dagger \phi\left( \phi^\dagger H +H^\dagger \phi\right)$ will spoil 
the accidental $Z_2$ symmetry.
\item The Lagrangian should of course be invariant under SU(2)$_L\times$ U(1)$_Y$ before electroweak symmetry breaking. 
So the existence of the Yukawa interaction $L\chi_i\phi_i$ for fermions $\chi_i$
and scalars $\phi_i$ implies 
\begin{equation}
d_{\phi_i} + d_{\chi_i} = 2\, n +1 \; , 
\end{equation}    
where $d_{\phi_i}$ and $d_{\chi_i}$ denote the dimension of $\phi_i$ and $\chi_i$, and $n$ is a positive integer.
Similarly a scalar trilinear coupling $\phi_i\phi_j H$ will result in 
\begin{equation}
d_{\phi_i} + d_{\phi_j} = 2\, n + 1\; .
\end{equation}
\end{enumerate}

We discuss the four possible topologies in turn:
For topology T1-i, any of the scalar field $\phi_i$ being a dark matter candidate will assign another scalar field $\phi_j$ with hypercharge $\pm\frac{1}{2}$
through trilinear scalar couplings $\phi_i \phi_j H$.
So according to condition 1 and 2, both $\phi_i$ and $\phi_j$ are odd-dimensional,
which violates condition 3. 
Alternatively $\chi_i$ can play the role of dark matter. 
This will also assign $\phi_i$ with hypercharge $\pm \frac{1}{2}$ through Yukawa interaction $L\chi_i \phi_i$.
Similarly both $\chi_i$ and $\phi_i$ are odd-dimensional according to condition 1 and 2, 
which contradicts with condition 3.
So topology T1-i does not provide any valid UV completion. 

We follow the same argument for topology T1-ii.
It suffices to consider either $\phi_1$ or $\chi_2$ to be a dark matter candidate as the diagram is symmetric. 
So the only possible assignment of hypercharges is $Y(\phi_1) =0$, $Y(\phi_2)=-\frac{1}{2}$, $Y(\chi_1)=\frac{1}{2}$ and $Y(\chi_2)=0$.
We immediately notice that $\phi_2$ has to be in an odd-dimensional representation according to condition 2. 
Therefore if $\phi_1$ is the dark matter candidate,  it must be an
odd-dimensional representation of SU(2)$_L$ 
and does not satisfy condition 3. 
Alternatively if $\chi_2$ is the dark matter candidate,  it should be in an
odd-dimensional representation and does not satisfy condition 3. 
So topology T1-ii is also ruled out. 

For topology T1-iii,  we find two minimal completions as shown in Tab.~\ref{tab:t1-iiimodel}, one for scalar dark matter and one for fermion dark matter.
Besides the exotic particles shown in the loop, another fermion $\bar{\psi}$ has to be introduced to cancel the anomaly
and to write down a mass term.

\begin{table}[b]
\centering
\begin{tabular}{c  ccc c }
\toprule
 Spin of DM  & $\chi$ & $\psi$ & $\bar{\psi}$ & $\phi$ \\
\midrule
0                     &$(5,0) $  &  $(6, \frac{1}{2}) $   & $(6,-\frac{1}{2}) $    & $(7,0) $ \\ 
 $\frac{1}{2}$   & $(5,0)$  &  $(4, \frac{1}{2})$   & $ (4,-\frac{1}{2})$    & $(5,0)$  \\ 
\bottomrule
\end{tabular}

\vspace{2ex}

\begin{minipage}{11cm}
	\caption{Matter content of both the scalar and fermionic dark matter model for
	T1-iii and their quantum numbers under 
SU(2)$_L\times$ U(1)$_Y$. Weyl notation is used.}
\label{tab:t1-iiimodel}
\end{minipage}
\end{table}

We turn to topology T3 in the end. If we take $\chi_i$ to be dark matter candidate, $\phi_i$ should have hypercharge $\frac{1}{2}$ and 
thus be an odd-dimensional representation according to condition 2, which is contradicting with condition 3. 
This failed attempt is exactly the model considered in Ref.~\cite{Cai:2011qr}. 
Another possibility is $\phi_i$ being the dark matter candidate.  Again we can just take $\phi_1$ due to the symmetry of the diagram.
The quantum numbers of the exotic particles of the minimal UV completion of T3 are
listed in Tab.~\ref{tab:t3model}. Non-minimal completions of T3 contain larger
SU(2) representations.
\begin{table}[tb]
\centering
\begin{tabular}{c  cc cc}
\toprule
 Spin of DM & $\phi_1$ & $\phi_2$ & $\chi_1$ & $\chi_2$ \\
\midrule
0 & $(7,0)$  &  $(5, 1)$   & $(6,\frac{1}{2})$    & $(6,-\frac{1}{2})$ \\ 
\bottomrule
\end{tabular}

\vspace{2ex}

\begin{minipage}{10cm}
	\caption{Matter content of the scalar dark matter model for T3 and their quantum
	numbers under SU(2)$_L\times$ U(1)$_Y$.
Weyl notation is used.}
\label{tab:t3model}
\end{minipage}
\end{table}

The last sanity check is to see if any renormalizable term invariant under the
SM gauge group breaks the accidental $Z_2$.
For the completions shown in Tab.~\ref{tab:t1-iiimodel}, the Yukawa coupling
$\psi\bar{\psi}\phi$ spoils $Z_2$~\cite{1603.04723} as well as the
cubic coupling $\phi^3$~\cite{Cirelli:2009uv,1603.01247v1}. Similarly $\chi_1\chi_2 \phi_1$ and
$\phi_1\phi_2^\dagger\phi_2$ for the model in
Tab.~\ref{tab:t3model} do the same~\cite{1603.04723}. Similar arguments
apply to higher representations.
Hence, there is no one-loop UV completion of the Weinberg operator in the
framework of minimal dark matter, which leads
to an exact accidental $Z_2$ symmetry. 
However, in the limit of vanishing renormalizable $Z_2$-breaking couplings, the symmetry of the
Lagrangian is enlarged and thus it is technically natural to have small
$Z_2$-breaking couplings~\cite{'tHooft:1979bh}. All quantum corrections to these couplings are proportional
to $Z_2$-breaking couplings.
With regards to minimal dark matter, the fermionic dark matter candidate is
preferred~\cite{Cirelli:2009uv}. We will consider the fermionic dark matter model in
Tab.~\ref{tab:t1-iiimodel} and perform a detailed
phenomenological study under the assumption that the $Z_2$-breaking couplings
are sufficiently small, such that the lifetime of $\chi$ is
longer than the age of the Universe and satisfies all indirect detection
constraints. Indirect detection constraints from photon, $\nu$, $e^+$ and $p^-$
searches are the strongest and generally require lifetimes $\tau_{DM}\gtrsim
10^{26}$ sec~\cite{Ibarra:2013zia,Rott:2014kfa,Ando:2015qda,Giesen:2015ufa}. The
maximum allowed size of the coupling can be estimated from the bound
given in Eq.~(13) in Ref.~\cite{1603.04723}
\begin{equation}
	\tau_\mathrm{DM}\lesssim 7.1\times 10^8 \left(\frac{10^{-2}}{\lambda}\right)^2
	\left(\frac{10^4 \mathrm{GeV}}{m_\mathrm{DM}}\right)^5
	\left(\frac{m}{10^9\mathrm{GeV}}\right)^4 \mathrm{sec}\;,
\end{equation}
where $\lambda$ denotes the $Z_2$-breaking Yukawa coupling and $m$ the mass of
the particles in the loop. Taking $m\sim m_\mathrm{DM}$, the bound on
$\tau_\mathrm{DM}$ can be translated to a bound on the $Z_2$-breaking coupling, 
$\lambda\lesssim 3\times 10^{-21}$. Despite the coupling being required to be
extremely small, it is technically natural and quantum corrections will not
induce larger $Z_2$-breaking couplings.
Thus the neutral component of $\chi$ constitutes a viable dark
matter candidate.
Note that besides the three models in Tabs.~\ref{tab:t1-iiimodel} and
\ref{tab:t3model} there are other one-loop radiative neutrino mass models with a
viable dark matter candidate including the original R$\nu$MDM model~\cite{Cai:2011qr}. 

\section{The Model}

\label{sec:model}
The R$\nu$MDM model with an approximate $Z_2$ symmetry, which we consider in
this work, contains a real scalar quintuplet $\phi$, 
two fermionic quadruplets $\psi$ and $\bar{\psi}$, which form a Dirac pair, and a fermionic quintuplet $\chi$. 
All even-dimensional SU(2) representations are
pseudo-real and all odd ones are real.
The higher-dimensional representations can be easily constructed using raising and lowering operators
and we list the ones relevant for this model in Appendix~\ref{app:su2}.     

The real scalar quintuplet $\phi$ can be decomposed into electrical charge eigenstates as 
$\phi=\left( \phi^{++}, \phi^+, \phi^0, \phi^-,\phi^{--}\right)^T$, 
where superscripts denote the electrical charges.
The neutral component field $\phi^0$ is a real scalar and  
the charged component fields $\left(\phi^+\right)^\dagger = \phi^-$,
$\left(\phi^{++}\right)^\dagger = \phi^{--} $ form two complex scalars.
Similarly for the fermions, the component fields are defined as
$\chi=\left(\chi^{++}, \chi^+, \chi^0, \chi^-, \chi^{--} \right)^T$,
$\psi=\left(\psi^{++}, \psi^+, \psi^0, \psi^- \right)^T$ and 
$\bar{\psi}=\left(\bar{\psi}^{+}, \bar{\psi}^0, \bar{\psi}^-, \bar{\psi}^{--} \right)^T$. 

Besides the SM part, the Lagrangian of this model consists of kinetic terms of the exotic particles 
and additional Yukawa interactions
\begin{align}
\mathcal{L} = \mathcal{L}^{SM}+  \mathcal{L}^{kin} + \mathcal{L}^{yuk}. 
\end{align}
We leave the details of the kinetic terms to Appendix~\ref{app:su2} as well.
From the discussion later in this section, we know that more than one copy of
$\phi$ is needed to explain both, the solar and the atmospheric, mass 
splittings.
Thus we attach a subscript as the family indices $\phi_i$.        
Then the Yukawa interactions are explicitly expressed in component fields as
\begin{align}
\label{eqn:yukawa}
\mathcal{L}^{yuk}&= Y^H H^\dagger \psi \chi  +\sum_{i,j} Y^L_{ij}  L_i  \psi \phi_j + h.c. \\
\nonumber
& = \frac{Y^H}{2} \left\{ H^- \left(2\chi^{++}\psi^- -\sqrt{3}\chi^+\psi^0+\sqrt{2}\chi^0\psi^+ -\chi^-\psi^{++} \right)\right.\\ 
&  \qquad  \quad  + \left.  (H^0)^\dagger \left(\chi^+\psi^- - \sqrt{2}\chi^0\psi^0 +\sqrt{3}\chi^-\psi^+ - 2\chi^{--}\psi^{++}\right)\right\} \label{eqn:yh} \\
\nonumber 
& + \frac{Y^L_{ij}}{2} \left\{e_i \left(2\phi^{++}_j\psi^- - \sqrt{3}\phi^{+}_j\psi^0 + \sqrt{2}\phi^{0}_j\psi^+ -\phi^{-}_j\psi^{++} \right) \right. \\
&  \qquad  \quad +  \left.\nu^i \left( -\phi^{+}_j \psi^- +
\sqrt{2}\phi^{0}_j\psi^0 -\sqrt{3}\phi^{-}_j \psi^+ + 2 \phi^{--}_j\psi^{++}
\right)\right\} + h.c.\;.
\label{eqn:yl}
\end{align}
Without loss of generality we can choose $Y^H$ to be real and positive using a
phase redefinition of $\psi$. Similarly three phases of $Y_{ij}^L$ can be
absorbed by a phase redefinition of $L_i$.
 
Apparently the Yukawa interaction in Eqn.~\eqref{eqn:yh} will induce mixing in the fermion sector.
The mass matrix for the neutral sector in basis $\Psi^0=\left(
\psi^0,\bar{\psi}^0, \chi^0 \right)$ is
\begin{align}
M_0=\left(
\begin{tabular}{ccc}
0 & $m_\psi$ & $\frac{1}{2} Y^H v$ \\
$m_\psi$ & 0 & 0 \\
$\frac{1}{2}Y^H v$ & 0 & $m_\chi$ 
\end{tabular}
\right)\; ,
\end{align}
where $v=246$ GeV is the vev of the Higgs boson.
Similarly the mass matrix for the singly charged fermions in the basis 
	$\left( \psi^+ , \bar{\psi}^+ , \chi^+, \psi^- , \bar{\psi}^- , \chi^-
	\right)$ reads
\begin{align}
M_1=\left(
\begin{tabular}{cc}
0 & $X_1$ \\
$X_1^T$ & 0  
\end{tabular}
\right)
\qquad
X_1 = \left(
\begin{tabular}{ccc}
0 & $-m_\psi$ & $-\frac{\sqrt{3}}{2\sqrt{2}} Y^H v$ \\
$-m_\psi$  & 0 & 0 \\
$-\frac{1}{2\sqrt{2}} Y^H v$ & 0 & $-m_\chi$ 
\end{tabular}
\right)\;.
\end{align}
The doubly-charged fermions also mix through the Yukawa interaction. In the basis 
of interaction eigenstates $\left(\psi^{++}, \chi^{++}, \bar{\psi}^{--}, \chi^{--}\right)$, the mass matrix reads
\begin{align}
M_2=\left(
\begin{tabular}{cc}
0 & $X_2$\\
$X_2^T$ & 0 
\end{tabular}\right)
\qquad
X_2=\left(
\begin{tabular}{cc}
$m_\psi$ & $\frac{1}{\sqrt{2}} Y^H v$ \\
0 & $m_\chi$
\end{tabular}
\right)   \; .
\end{align}
The diagonalization of these mass matrices can be done using the Takagi
factorization and singular value decomposition, respectively. The details of which are collected in Appendix~\ref{app:diagonalization}.

The neutral component of $\chi$ is the dark matter candidate of this model. 
To saturate the dark matter density observed by Planck~\cite{Ade:2015xua}, the
dark matter mass should be\cite{Cirelli:2015bda}   
\begin{equation}
	m_\chi = \left(9.4\pm0.47\right) \;  \rm{TeV} ,
\end{equation} 
where Sommerfeld effects have been taken into account~\cite{Cirelli:2009uv}.
Other dark sector particles have to be heavier than $m_\chi$, which
makes the mixing between $\chi$ and $\psi$, roughly given by $Y^H v/m_{\chi,\psi}$, negligible. 

The neutrino mass after electroweak symmetry breaking is defined by 
$\mathcal{L}=-\frac12 m_\nu \nu\nu$ and thus
\begin{equation}
(m_\nu)_{ij}=\frac{(Y^H v)^2 m_\chi}{256\pi^2 m_\psi^2}
\sum_k Y_{ik}^L Y_{jk}^L\; 
g\left(\frac{m_{\phi_k}^2}{m_\psi^2},\frac{m_\chi^2}{m_\psi^2}\right)
\end{equation}
with the factor  
\begin{equation}
	g(\eta_1,\eta_2) = \frac{1}{(1-\eta_1)(1-\eta_2)}
	+\frac{1}{\eta_1-\eta_2}\left(\frac{\eta_1^2\ln\eta_1}{(1-\eta_1)^2}
	-\frac{\eta_2^2\ln\eta_2}{(1-\eta_2)^2}\right)\; .
\end{equation}
The calculation is done neglecting mass mixing among the fermions which could only give rise to 
a tiny correction proportional to $Y^H v/m_{\chi,\psi}$. 
The neutrino mass matrix $m_\nu$ is full rank if and only if $n_\phi \geq 3$. 
The Yukawa couplings $Y^L$ can be expressed as
\begin{equation}
Y^L_{ij} = \frac{16 \pi \;  m_\psi}{Y^H \; v \sqrt{m_\chi}}\sum_{k,l} \left(V_\nu^*\right)_{ik} \left(\hat{m}_\nu^\frac{1}{2} \right)_{kl}\mathcal{O}_{lj}
\;  g^{-\frac{1}{2}}_j 
\end{equation}
using a Casas-Ibarra-type parametrisation~\cite{Casas:2001sr}, 
where $\mathcal{O}$ is a general complex orthogonal matrix and $U$ diagonalises the neutrino mass matrix with $\hat{m}_\nu = V_\nu^T m_\nu V_\nu$.  

To keep minimality, we will only introduce two copies of $\phi$, which suffices
to explain the solar and atmospheric mass splittings. One of the neutrinos
remains massless.
Both $Y^L$ and $\mathcal{O}$ are $3\times 2$ matrices. The Yukawa couplings
$Y^L$ are entirely determined by the solar and atmospheric mass squared differences and
the leptonic mixing parameters. The only relevant undetermined parameter is a complex
angle $\theta$ parameterising the $3\times 2$ orthogonal matrix $\mathcal{O}$. The
two elements of the orthogonal matrix $\mathcal{O}$ associated with the massless
neutrino do not enter the Yukawa coupling $Y^L$. Neutrino
mass fixes the product of the Yukawa couplings $Y^L_{ij} Y^H$. 
We will study the phenomenology induced by these Yukawa couplings and work out the experimental constraints on
the model parameter space in the following sections. 

\section{Constraints from Higgs physics and flavor physics} 
\label{sec:flavor}
The Yukawa interactions introduced in Eqn.~\eqref{eqn:yukawa} modify the SM in two aspects:
The coupling between the SM Higgs and the exotic particles will change the decay
width of the Higgs;
Lepton family number is also violated and lead to lepton flavor-changing rare processes.
Both impose constraints on the parameter space of the model.
Since exotic particles introduced are not charged under $SU(3)_c$, there will be no new contribution to processes such as
meson mixings and  $b\to s$ transition.  
Derivation of the constraints involves calculation of one-loop Feynman diagrams,
which is assisted by the Mathematica packages
FeynRules~\cite{Christensen:2008py}, FeynArts~\cite{Hahn:2000kx},
FormCalc~\cite{Hahn:1998yk}, and ANT~\cite{Angel:2013hla}

\subsection{Constraints from the Higgs}
\begin{figure}[t!]
 \centering
   \begin{tikzpicture}[node distance=1cm and 1cm]
        \coordinate[ label=left:$h$] (v1);
        \coordinate[vertex, right=of v1] (v2);
        \coordinate[cross, right=of v2, xshift=1cm] (v3);
        \coordinate[vertex, above=of v3] (v4);
        \coordinate[vertex, below=of v3] (v5);
        \coordinate[above right=of v4, label=right:$\gamma(Z)$] (v6);
        \coordinate[below right=of v5, label=right:$\gamma$] (v7);
        \draw[fermion] (v2) -- (v4) node[midway, above=0.1cm] {$\chi$};
        \draw[fermion] (v4) -- (v3) node[midway, right=0.1cm] {$\chi$};
        \draw[fermion] (v2) -- (v5) node[midway, below=0.1cm] {$\psi$};
        \draw[fermion] (v5) -- (v3) node[midway, right=0.1cm] {$\psi$};
        \draw[scalarnoarrow] (v1) -- (v2) ;
        \draw[photon] (v4) -- (v6) ;
        \draw[photon] (v5) -- (v7) ;
        \draw[fermionnoarrow] (v5) -- (v4) ;
   \end{tikzpicture}

   \begin{minipage}{10cm}
	   \caption{Higgs decays to two bosons at one loop with exotic fermions. There are similar diagrams with the mass insertion 
            on the other two edges of the fermion triangle.  }
	  \label{fig:hdiboson}
  \end{minipage}
\end{figure}
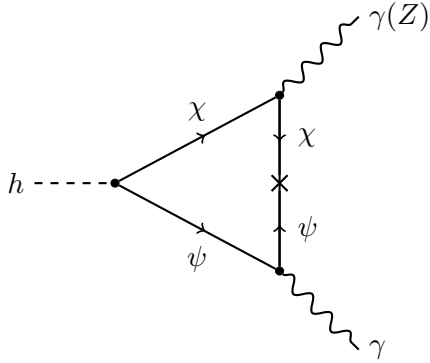

\begin{figure}[tb]
\centering
\includegraphics[width=0.5\textwidth]{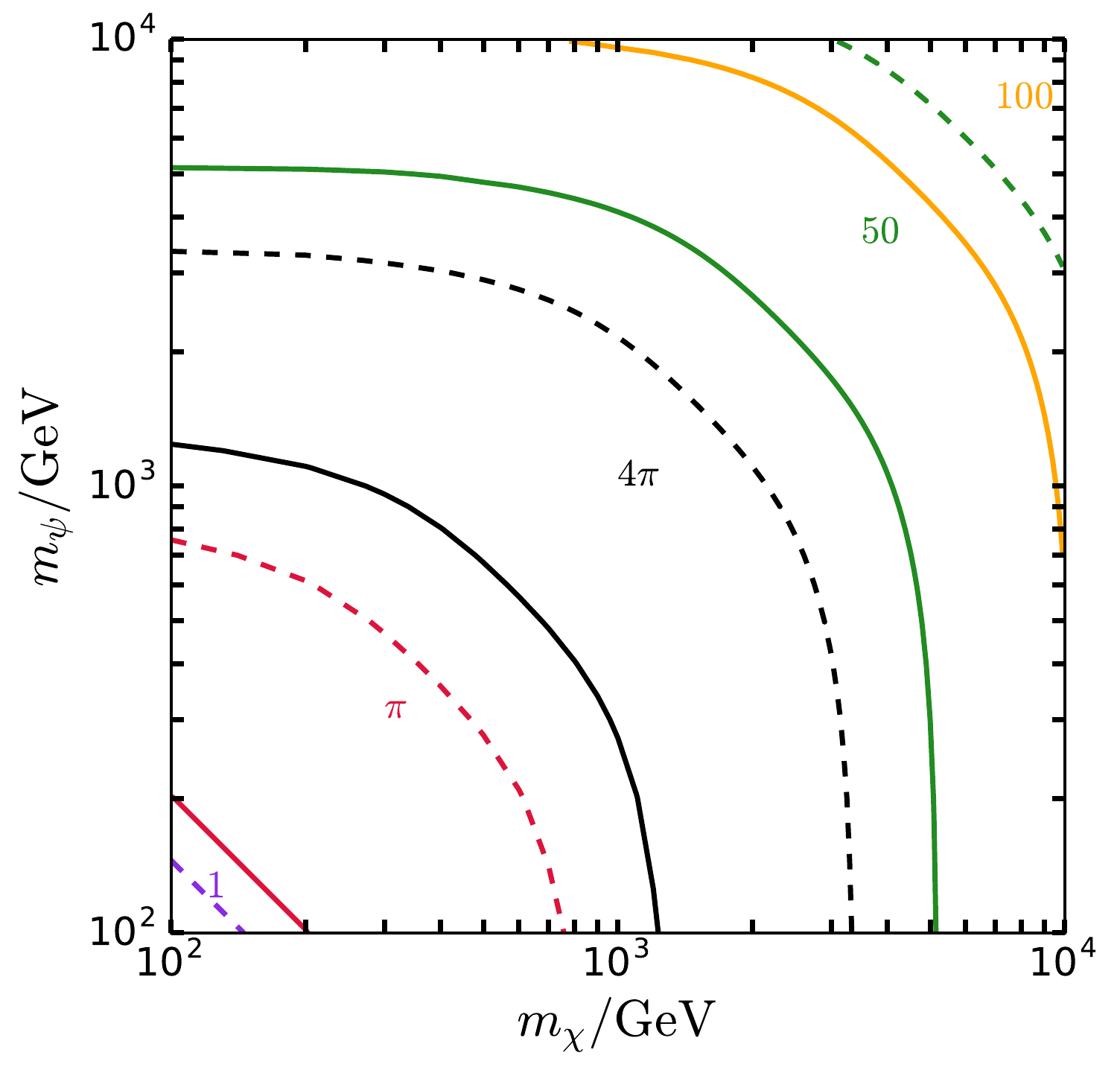}
\caption{Contour plot of the maximal $Y^H$ allowed by the measurement of the signal strength of $h\to \gamma\gamma$.
The solid contours denotes the current limit and the dashed from the future 3000 $fb^{-1}$ LHC,
where purple, red, black, green and yellow lines correspond to $1$, $\pi$, $4\pi$, $50$ and $100$. }
\label{fig:ymax}
\end{figure}

The Yukawa coupling $Y^H$ can modify the behavior of the Higgs boson.
As all exotic particles are colorless, the production of the Higgs at the LHC is untouched.    
The decay from the Higgs to the exotic particles is also forbidden kinematically.  
However, the exotic particles can contribute to the decay of the Higgs at loop level such as $h\to \gamma\gamma$
or $h\to Z\gamma$ as depicted in Fig.~\ref{fig:hdiboson}.
LHC has measured the production of the Higgs boson in the diphoton channels. Among them
the diphoton measurement is the most accurate one, which we will use to derive our constraint.       
The Higgs signal strength in the diphoton channel is defined as  
\begin{align}
\mu = \frac{\Gamma_{h\to \gamma\gamma}^{\rm R\nu MDM}}{\Gamma_{h \to \gamma\gamma}^{\rm SM}} \; ,
\end{align}
with the decay width of Higgs to diphoton in this model  
\begin{align}
\Gamma_{h\to \gamma\gamma}^{\rm R\nu MDM} & = \frac{G_F \alpha^2 m_h^3}{128 \sqrt{2}\pi^3} 
            \left|F_1(\frac{4 m_W^2}{m_h^2}) + \sum_f N_c Q_f^2 F_\frac{1}{2}(\frac{4 m_f^2}{m_h^2}) 
        + \;  \frac{10\times \left| Y^H \right|^2 m_t v}{m_\chi^2} I(\frac{m_\psi^2}{m_\chi^2})\right|^2 \; , 
\label{eqn:gammaBSM}
\end{align}
where the first two terms are the well-known results for the contributions from $W$ boson and the SM fermions,
and the last term is the new contribution from the Yukawa interaction $H^\dagger\psi\chi$.
$\Gamma_{h\to \gamma\gamma}^{\rm SM}$ can be achieved by simply setting $Y^H$ to
be zero in Eqn.~\eqref{eqn:gammaBSM}.   
The dimensionless loop factors in Eqn.~\eqref{eqn:gammaBSM} are 
\begin{align}
F_1(\tau ) & = 2\tau + 3 \tau + 3 \tau (2-\tau) f(\tau)\;, &
	F_\frac{1}{2} (\tau) & = - 2 \tau (1 + (1-\tau)f(\tau))\;, \nonumber\\
I(\eta)& =\frac{(\eta +1) \left(\eta ^3+9 \eta ^2-9 \eta - 1 -6 \eta (\eta +1)  \ln \eta \right)}{6 (\eta -1)^5} \; , &
\end{align}
where $f(\tau) = \sin^{-1} (\sqrt{1/\tau})$ and $I(\eta)$ has been obtained in
the limit $m_\chi, m_\psi \gg  m_h$.
We derive the limit on $Y^H$ for specific fermion masses from $\mu=1.17_{-0.17}^{+0.19}$~\cite{Agashe:2014kda}.
In Fig.~\ref{fig:ymax} we plot the contours of the maximal $Y^H$ allowed by the measurements at the LHC in solid lines.
As the exotic particles have masses of $9.4$ TeV to saturate the relic density, 
the current experiments impose literally no constraints on $Y^H$.
We also project the future constraints on $Y^H$ at 14 TeV LHC with 3000 $fb^{-1}$ dataset.
An optimistic estimate of 5\% sensitivity~\cite{ATL-PHYS-PUB-2013-014} 
is used to extract the limits shown in the dashed contours also in Fig.~\ref{fig:ymax}.
However, the heavy masses of the exotic particles make it impossible to place any meaningful constraints 
on $Y^H$ even at the high luminosity (HL) -LHC.   

\subsection{Lepton Flavor Violating Processes}
The Yukawa interaction in Eqn.~\eqref{eqn:yl} will induce lepton flavor
violating (LFV) processes. 
Several LFV processes,
such as $\mu^-\to e^-\gamma$, $\mu\to eee$ and $\mu$-$e$ conversion in nuclei, have been probed
with extremely high sensitivity, but no signal has been found. 
The experimental results will surely place strong constraints on the model
parameters. We will study the most well measured and thus most stringent ones in
this section. The implications of these constraints will be discussed in
Sec.~\ref{sec:other}.

\subsubsection{$l_i\to l_j \gamma$}
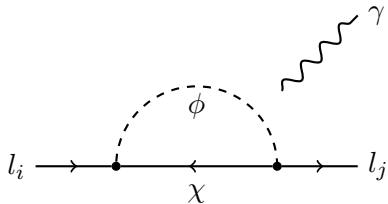
\begin{figure}[tb]
 \centering
   \begin{tikzpicture}[node distance=1cm and 1cm]
        \coordinate[ label=left:$l_i$] (v1);
        \coordinate[vertex, right=of v1] (v2);
        \coordinate[vertex, right=of v2, xshift=1cm] (v3);
        \coordinate[right=of v3, label=right:$l_j$] (v4);
        \coordinate[above left=of v4] (v5);
        \coordinate[above right=of v5, label=right:$\gamma$] (v6);
        \draw[fermion] (v1) -- (v2) ;
        \draw[fermion] (v3) -- (v4) ;
        \draw[fermion] (v3) -- (v2) node[midway, below=0.1cm] {$\chi$};
        \semiloop[scalarnoarrow]{v2}{v3}{0};
	\node at ($(v2)!0.5!(v3)+(0,0.8)$) {$\phi$};
        \draw[photon] (v5) -- (v6);
   \end{tikzpicture}

   \begin{minipage}{10cm}
	   \caption{Feynman diagram for $l_i\to l_j \gamma$ where the emitted photon can be attached to any charged particles in the diagram.  }
\label{fig:muegamma}
   \end{minipage}
\end{figure}

We consider the process of the form $l_i\to l_j\gamma$ first and follow the convention in Ref.~\cite{Lavoura:2003xp}.
This type of decay is rare compared with the
	dominant tree-level decay via a virtual $W$ boson.
The amplitude of such process, depicted in Fig.~\ref{fig:muegamma}, reads     
\begin{align}
\mathcal{M}(l_i\to l_j \gamma) = e \; \epsilon_\mu^*\bar{u}(p_j) i\sigma^{\mu\nu}q_\nu(\sigma_{Lij} P_L +\sigma_{Rij} P_R) u(p_i)\ ,
\end{align}
where $e$ is the magnitude of electron charge and $P_{L,R}\equiv \frac{1}{2}\left(1\mp \gamma_5\right)$ 
are the projection operators.
The coefficients $\sigma_{Lij,Rij}$ can be written as
\begin{align}
&\sigma_{Lij}= \frac{m_{l_j}}{16\pi^2} \sum_k
	\frac{Y^L_{ik}Y^{L*}_{jk}}{m_{\phi_k}^2}
	F(\frac{m_\psi^2}{m_{\phi_k}^2})\;,
&\sigma_{Rij}= \frac{m_{l_i}}{16\pi^2} \sum_k \frac{Y^L_{ik}Y^{L*}_{jk}}{m_{\phi_k}^2} F(\frac{m_\psi^2}{m_{\phi_k}^2}) \; ,
\label{eqn:sigmalr} 
\end{align}
where $F(\eta)$ is defined as
\begin{align}
F(\eta)=\frac{5\left(1-6\eta + 3\eta^2 +2 \eta^3 - 6 \eta^2 \ln\eta \right)}{24\left(\eta-1\right)^4} \; .
\end{align}
The partial width then can be easily calculated with
\begin{equation}
\Gamma(l_i\to l_j \gamma) = \frac{\left(m_{l_i}^2-m_{l_j}^2\right)^3 e^2 \left(|\sigma_L|^2+|\sigma_R|^2\right)}{16\pi m_{l_i}^3}
\end{equation}
In the minimal variant of the model with two scalar quintuplets only, the
relative rates are fixed and they are of similar order of magnitude. The current experimental limits are given by Br$(\mu\to e\gamma)<5.7\times 10^{-13}$~\cite{Adam:2013mnn}, 
Br$(\tau\to e\gamma)<3.3\times 10^{-8}$ and Br$(\tau\to \mu \gamma)<4.4\times
10^{-8}$~\cite{Agashe:2014kda}. Thus the most constraining limit is from $\mu\to
e\gamma$, which we will discuss in Sec.~\ref{sec:other}. The proposed upgrade of
MEG will improve the future sensitivity of $\mu\to e\gamma$ down to Br$(\mu \to
e\gamma)\sim6\times 10^{-14}$~\cite{Baldini:2013ke}.    

\subsubsection{Anomalous Magnetic Moment}
The lepton anomalous magnetic moment can also be expressed with the terms in
Eqn.~\eqref{eqn:sigmalr}
\begin{equation}
\Delta a_i = 2\; e \; m_{l_i} \left( \sigma_{Lii}+\sigma_{Rii}\right) 
= \frac{e}{4\pi^2}\sum_k \left|Y_{ik}^L\right|^2\frac{m_{l_i}^2}{m_{\phi_k}^2} F(\frac{m_\psi^2}{m_{\phi_k}^2}) \; .
\end{equation}
The anomalous magnetic moment of the muon has been measured to a very high
precision $\Delta a_\mu=0.0011659209\pm0.0000000006$ \cite{Bennett:2006fi}.
Given the heavy scalar quintuplet masses, the correction to the anomalous
magnetic moment of the muon is well below the experimental precision and thus
it does not impose any constraint.

\subsubsection{$\mu\to eee$}
\begin{figure}[t!]
 \centering
    \begin{subfigure}[t]{0.45\textwidth}
        \centering
        \begin{tikzpicture}[node distance=1cm and 1cm]
          \coordinate[label=left:$\mu^-$] (v1); 
          \coordinate[right=of v1] (v2);
          \coordinate[right=of v2, xshift=1cm] (v3);
          \coordinate[right=of v3, label=right:$e^-$] (v4);
          \coordinate[below left=of v3, yshift=0.5cm] (v5);
          \coordinate[below right=of v5] (v6);
          \coordinate[above right=of v6, yshift=-0.3cm, label=right:$e^-$] (v7); 
          \coordinate[below right=of v6, yshift=0.3cm, label=right:$e^+$] (v8); 
          \semiloop[scalarnoarrow]{v2}{v3}{0};
	  \node at ($(v2)!0.5!(v3)+(0,1.3)$) {$\phi$};
          \draw[fermion] (v1)--(v2);
          \draw[fermion] (v3)--(v2) node[midway,above=0.1cm] {$\psi$};
          \draw[fermion] (v3)--(v4);
          \draw[photon]  (v5)--(v6) node[midway, left=0.3cm] {$\gamma(Z)$};
          \draw[fermion] (v6) -- (v7);
          \draw[fermion] (v8) -- (v6);
        \end{tikzpicture}
    \end{subfigure} 
    \begin{subfigure}[t]{0.45\textwidth}
        \centering
        \begin{tikzpicture}[node distance=1cm and 1cm]
          \coordinate[label=left:$\mu^-$] (v1);
          \coordinate[right=of v1] (v2);
          \coordinate[right=of v2, xshift=1cm] (v3);
          \coordinate[right=of v3, label=right:$e^-$] (v4);
          \coordinate[above right=of v2] (v5);
          \coordinate[below right=of v2] (v6);
          \coordinate[ above right=of v5, label=right:$e^-$] (v7);
          \coordinate[ below right=of v6, label=right:$e^+$] (v8);
          \draw[fermion] (v1) -- (v2);
          \draw[fermion] (v5) -- (v2) node[midway,above=0.1cm] {$\psi$};
          \draw[fermion] (v5) -- (v7);
          \draw[fermion] (v8) -- (v6);
          \draw[fermion] (v3) -- (v4);
          \draw[fermion] (v3) -- (v6) node[midway, right=0.1cm] {$\psi$};
          \draw[scalarnoarrow] (v5) -- (v3) node[midway,right=0.1cm] {$\phi$};
          \draw[scalarnoarrow] (v2) -- (v6) node[midway,left=0.1cm] {$\phi$};
        \end{tikzpicture}
    \end{subfigure} 
\caption{Feynman diagrams for $\mu^-\to e^-e^+e^-$:  $\gamma$- or $Z$-penguins on the left and the box diagram on the right.}
\label{fig:mueee}
\end{figure}
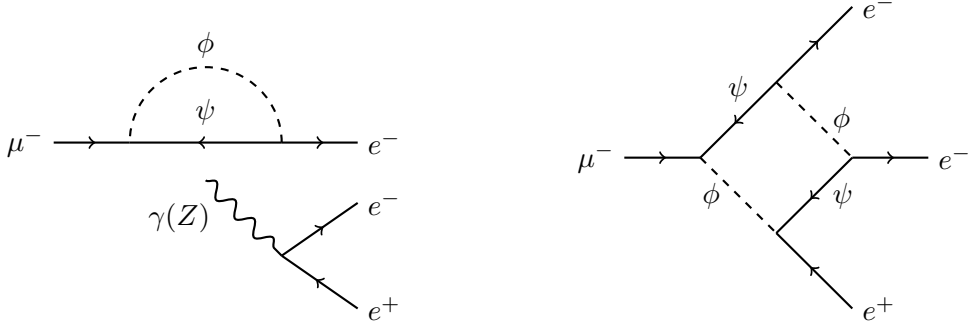

Another well-measured LFV process is $\mu\to eee$, which this model will have several different contributions
including $\gamma$-penguin, $Z$-penguin, Higgs-penguin, and box diagrams. The
contribution from the Higgs-penguin is proportional to 
the electron mass and negligible. We show the relevant Feynman diagrams for this process in Fig.~\ref{fig:mueee}.
The amplitude of the $\gamma$-penguin can be written in the form of    
\begin{align}
\nonumber
\mathcal{M}_{\gamma} &= \bar{u}(p_1)\left(q^2 \gamma_\mu (A_1^L P_L +A_1^R P_R)
+ i m_\mu \sigma_{\mu\nu}q^\nu\left( A_2^L P_L +A_2^R P_R \right)
\right) u(p) \\
&\times \frac{e^2}{q^2} \bar{u}(p_2)\gamma^\mu v(p_3) -\left( p_1\leftrightarrow p_2 \right) \; , 
\end{align}
where $p$ is the momentum of the initial muon and $p_{1,2,3}$ are the momenta of the two electrons and the positron in the final state.  
The loop functions are given by
\begin{align}
A_1^L &=  \frac{1}{16\pi^2} \sum_k \frac{ Y^L_{2k}Y^{L*}_{1k}}{m_{\phi_k^2}}
	G(\frac{m_\psi^2}{m_{\phi_k}^2})\;, & 
 G(\eta)&=\frac{5\left(2 - 9 \eta + 18\eta^2 -11\eta^3 + 6\eta^3
 \ln\eta\right)}{72\left(\eta-1\right)^4}\;,  \\
A_1^R &= 0\;, &
 A_2^{L,R} &= \frac{\sigma_{L21,R21}}{m_\mu},
\end{align}
where we have set the electron mass and the external momenta to zero.   
Similarly the contribution from the $Z$-penguin is  
\begin{align}
\mathcal{M}_Z&=\frac{1}{m_Z^2}\bar{u}(p_1) \gamma_\mu \left( F_L P_L+F_R P_R
	\right) u(p) \times \bar{u}(p_2) \gamma^\mu \left(Z_L^e P_L + Z_R^e
	P_R\right) v(p_3) - (p_1\leftrightarrow p_2)\;, \\ 
F_L&= \sum_k \frac{1}{16\pi^2}  \frac{  \; e \; Y^L_{2k}Y^{L*}_{1k}}{
\sin\theta_W \cos\theta_W} H(\frac{m_\psi^2}{m_{\phi_k}^2}) \;,
\quad F_R=0\;, 
\quad  H(\eta)=\frac{5\; \eta \left(1-\eta+\ln\eta\right)}{ 4\left(\eta-1\right)^2} \; ,
\end{align}
where $Z_{L,R}^e$ is the weak charge of the left- or right-handed electron.
The weak charge of any fermion $f$ is defined as   
\begin{equation}
 Z_{L,R}^f = \frac{e}{\sin\theta_W\cos\theta_W}\left(T_3^{f_L,f_R} -Q_f\sin\theta_W^2\right) 
\end{equation}
with $T_3^{f_L,f_R}$ being the isospin of $f_{L,R}$ and $Q_f$ the electrical charge of $f$. 
Finally the leading order contribution from the box diagram can be written as
\begin{align}
\mathcal{M}_{Box} = e^2 B_1^L \left[ \bar{u} (p_1) \gamma^\mu P_L u (p)\right] 
  \left[ \bar{u}(p_2) \gamma_\mu P_L v(p_3)\right] - (p_1\leftrightarrow p_2) \; ,
\end{align}
where we have neglected subdominant contributions further suppressed by the heavy mass scale 
and the loop function $B_1^L$ is  
\begin{align}
B_1^L= \sum_{i,j} \frac{1}{16\pi^2}\frac{5 Y^L_{2m} Y^{L*}_{1n} Y^{L*}_{1m} Y^L_{1n}}{4} 
D_{00}\left[m_\psi^2,m_\psi^2,m_{\phi_i}^2, m_{\phi_j}^2 \right] \; . 
\end{align}
The analytic expression of the four-point function $D_{00}$ can be found in the Appendix of Ref.~\cite{Angel:2013hla}. 
Among the various contributions, only the $Z$-penguin is suppressed by the $Z$
boson mass, while the others are all suppressed by the masses of the exotic particles. 
With the amplitude we can easily write down the decay width~\cite{Hisano:1995cp,Hisano:1998fj,Arganda:2005ji},
\begin{align}
\nonumber
\Gamma(\mu^-\to e^-e^+e^-) & =\frac{e^4 m_\mu^5}{512\pi^3}  \times  \left[ \left|A_1^L\right|^2  +  \left(\left|A_2^L\right|^2 + \left|A_2^R\right|^2\right)\left( \frac{16}{3}\ln\frac{m_\mu}{m_e}-\frac{22}{3}\right)  \right.\\
\nonumber
& + \frac{1}{6}\left|B_1^L\right|^2  +  \frac{2}{3}  \left|F_{LL}\right|^2 + \frac{1}{3} \left|F_{LR}\right|^2   - 2 \left(A_1^L A_2^{R*} + h.c. \right) \\
\nonumber
&   \left. + \frac{1}{3} \left( A_1^L B_1^{L*} - 2 A_2^R B_1^{L*}+ h.c.\right) + \frac{1}{3}\left(  2 A_1^L F_{LL}^* +  A_1^L F_{LR}^*  h.c.\right) \right. \\
& \left.  +  \frac{1}{3} \left(B_1^L F_{LL}^*\right) - \frac{2}{3}\left(2 A_2^R F_{LL}^* + A_2^R F_{LR}^* + h.c. \right)\right] \; , 
\label{eqn:brmueee}
\end{align} 
with the factors $F_{LL,LR} = F_L Z_{L,R}^e/(e\; m_Z)^2$, where we have omitted
all the vanishing or next-to-leading order contributions.
The heavy exotic
particles in this model imply that the $Z$-penguin contribution dominates over
the contributions of the $\gamma$-penguin and the box diagrams. 
The branching ratio for this rare decay channel can be approximately obtained by
dividing Eqn.~\eqref{eqn:brmueee} by the decay width of $\mu^-\to e^-\bar{\nu}_e \nu_\mu$. 
The current experimental limit is given by Br$(\mu\to
eee)<10^{-12}$~\cite{Bellgardt:1987du} and there is a proposal for an
experiment with an substantially increased sensitivity down to      
 Br$(\mu\to
 eee)<10^{-16}$~\cite{Blondel:2013ia}.

\subsubsection{$\mu$-$e$ Conversion in Nuclei}
The last LFV process we consider is $\mu$-$e$ conversion in nuclei. There are two types of contributions:
the long-range interactions determined by the electromagnetic dipole, and the short-range interactions from the penguin
diagrams as shown in Fig.~\ref{fig:mueee} with the final electron pair replaced by a quark pair. For this model, there is no contribution from the box diagrams because no colored exotic states are introduced.
Therefore the effective Lagrangian can be expressed as 
\begin{align}
\mathcal{L}_{eff}  = &-\frac{1}{2}m_\mu \left( A_2^L \bar{e} \sigma^{\mu\nu} P_L \mu  F_{\mu\nu}+ A_2^R \bar{e} \sigma^{\mu\nu} P_R \mu F_{\mu\nu} + h.c. \right)  \nonumber \\
      &  - \sum_{q=u,d,s}\left[ \left(g_{LV(q)} \bar{e}\gamma^\mu P_L \mu \right)\bar{q}\gamma_\mu q
          + \left( g_{LA(q)} \bar{e}\gamma^\mu P_L \mu \right) \bar{q}\gamma_\mu\gamma_5q + h.c.\right] \; ,
\label{eqn:mueconversion}
\end{align}         
where the first and second lines denote the long- and short-range interactions. 
The Wilson coefficients $g_{LV(q)}$ only receive a contribution from the $\gamma$-penguin, while $g_{LA(q)}$ from both $\gamma$- and $Z$-penguins. 
They can be written as 
\begin{align}
g_{LV(q)}^\gamma &=\frac{e^2\; Q_{q}}{16\pi^2}\sum_i
	\frac{Y^L_{2i}Y^{L*}_{1i}}{m_{\phi_i}^2}G(\frac{m_\psi^2}{m_{\phi_i}^2})\;, \\
g_{LV(q),LA(q)}^Z & = - \frac{e}{16\pi^2} \frac{\pm Z_L^{q}+ Z_R^{q}}{\sin\theta_W\cos\theta_W}\sum_i 
      \frac{Y^L_{2i}Y^{L*}_{1i}}{ m_Z^2} H(\frac{m_\psi^2}{m_{\phi_i}^2})  \; ,
\end{align} 
where $Q_f$ is the electrical charge of the quarks. Similar to $\mu\to
eee$ the $Z$-penguin contribution dominates over the $\gamma$-penguin.
Coherent processes generally dominate over incoherent processes, if the final
state nucleus is the same as the initial state nucleus. Thus we will focus on
coherent contributions to $\mu$-$e$ conversion and neglect any incoherent contribution.
So the conversion rate is~\cite{Kitano:2002mt} 
\begin{align}
\omega_{\rm conv} = 4 \left| \frac{1}{8} A_2^{R*} D + \tilde{g}_{LV}^{(p)} V^{(p)} + \tilde{g}_{LV}^{(n)} V^{(n)}\right|^2
                  + 4 \left| \frac{1}{8} A_2^{L*} D\right|^2 \; ,
\label{eqn:muerate}
\end{align}     
where $\tilde{g}_{LV}^{(p,n)}$ are the coefficients of the vector interactions with protons and neutrons defined as 
$\tilde{g}_{LV}^{(p)} = 2 g_{LV(u)}+g_{LV(d)}$ and $\tilde{g}_{LV}^{(n)} =
g_{LV(u)}+ 2 g_{LV(d)}$. We use the values in Tab.~1 of
Ref.~\cite{Kitano:2002mt} for the overlap integrals $D$, 
$V^{(p)}$ and  $V^{(n)}$.
The branching ratio of $\mu$-$e$ conversion is defined as  the ratio between the conversion rate and the capture rate 
\begin{align}
	{\rm Br}(\mu N \to e N) \equiv \frac{\omega_{conv}}{\omega_{capt}} \; ,  
\end{align}
where the rate $\omega_{capt}$ takes the value $13.07\times 10^6 s^{-1}$, $2.59\times 10^6 s^{-1}$
and $0.7054\times 10^{6} s^{-1}$ in Au, Ti and Al~\cite{Kitano:2002mt}. 
The current bounds on the branching ratio are  Br$(\mu {\rm Au}\to e {\rm Au}) < 7\times 10^{-13}$
and Br$(\mu {\rm Ti} \to e {\rm Ti}) <4.3\times 10^{-12}$~\cite{Agashe:2014kda}. 
There are good prospects to increase the future sensitivities for Al and Ti to Br$(\mu {\rm Al}\to e {\rm Al}) \lesssim 10^{-16}$
~\cite{Hungerford:2009zz,Cui:2009zz,Carey:2008zz,Kurup:2011zza,Kutschke:2011ux} and 
Br$(\mu {\rm Ti}\to e {\rm Ti}) \lesssim 10^{-18}$~\cite{Hungerford:2009zz,
Cui:2009zz}, respectively.


\section{Naturalness}
\label{sec:naturalness}

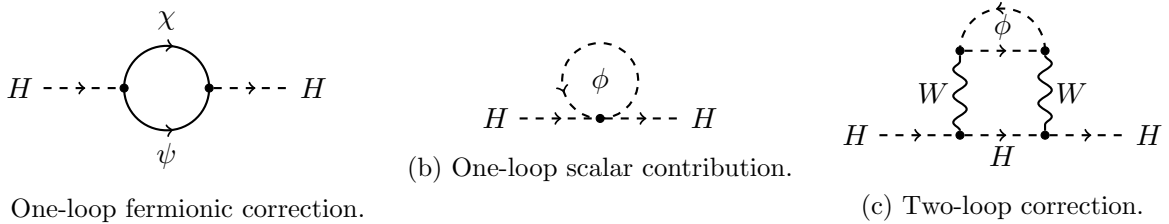
\begin{figure}[tb]\centering
	\begin{subfigure}{0.34\linewidth} \centering
		\begin{tikzpicture}[node distance=1cm and 1cm]
        \coordinate[ label=left:$H$] (v1);
        \coordinate[vertex, right=of v1] (v2);
        \coordinate[vertex, right=of v2] (v3);
        \coordinate[ right=of v3, label=right:$H$] (v4);
        \draw[scalar] (v1) -- (v2) ;
        \draw[scalar] (v3) -- (v4) ;
        \upsemiloop[fermion]{v2}{v3}{180};
	\node at ($(v2)!0.5!(v3)+(0,-0.9)$) {$\psi$};
	\lowsemiloop[fermion]{v2}{v3}{180};
	\node at ($(v2)!0.5!(v3)+(0,0.9)$) {$\chi$};
        \end{tikzpicture}
	\caption{One-loop fermionic correction.}\label{fig:OneLoopHiggsBilinear}
\end{subfigure}
\begin{subfigure}{0.34\linewidth}\centering
		\begin{tikzpicture}[node distance=1cm and 1cm]
        \coordinate[ label=left:$H$] (vL);
        \coordinate[vertex, right=of vL] (v);
        \coordinate[ below=of v] (gv);
        \coordinate[ right=of v, label=right:$H$] (vR);
	\draw[scalar] (vL) -- (v);
	\draw[scalar] (v) -- (vR);
	\node (phi) at ($(v)+(0,0.5)$) {$\phi$}; 
		\draw[scalar] (phi) circle [radius=0.5cm];
        \end{tikzpicture}
	\caption{One-loop scalar contribution.}
	\label{fig:OneLoopPhi}
\end{subfigure}
\begin{subfigure}{0.29\linewidth}\centering
	\begin{tikzpicture}[node distance=1cm and 1cm]
        \coordinate[ label=left:$H$] (vL);
        \coordinate[vertex, right=of vL] (v1);
        \coordinate[vertex, right=of v1] (v2);
	\coordinate[vertex, above=of v1] (v3);
	\coordinate[vertex, above=of v2] (v4);
        \coordinate[ right=of v2, label=right:$H$] (vR);
	\draw[photon] (v1) -- (v3) node[midway,left] {$W$};
	\draw[photon] (v4) -- (v2) node[midway,right] {$W$};
	\draw[scalar] (v1) -- (v2) node[midway,below] {$H$};
	\draw[scalar] (vL) -- (v1);
	\draw[scalar] (v2) -- (vR);
	\semiloop[scalar]{v3}{v4}{0};
	\draw[scalar] (v3) -- (v4) node[midway,above]{$\phi$};
        \end{tikzpicture}

	\caption{Two-loop correction.}\label{fig:TwoLoopHiggsBilinear}
\end{subfigure}
\caption{Corrections to Higgs bilinear from new particles in the theory.}
\end{figure}

The newly introduced particles lead to corrections to the Higgs effective 
potential. Defining the tree-level Higgs potential as
\begin{equation}
V(H) = - \mu_H^2 H^\dagger H +\lambda (H^\dagger H)^2
\end{equation}
with the Higgs vev $\braket{H}=v/\sqrt{2}$ and the Higgs mass $m_h^2$, the
minimization conditions give
\begin{equation}
	v^2=\frac{\mu_H^2}{\lambda}\;, \quad \mathrm{and}\quad m_h^2 =
	4\mu_H^2 = 4\lambda v^2\;.
\end{equation}
In order to estimate the corrections of the new particles on electroweak scale
naturalness, we calculate the corrections to the Higgs bilinear in
the scalar potential. The correction to the quartic term is dimensionless and
is not quadratically enhanced by the heavy mass scale compared to the
electroweak scale. We use dimensional regularization with
the modified minimal subtraction ($\overline{MS}$) renormalization scheme to calculate the
one-loop correction given in Fig.~\ref{fig:OneLoopHiggsBilinear}.
After canceling the divergent part with the counterterm, there is a finite
contribution to the effective bilinear term of the Higgs
$\mu_{H,eff}^2=\mu_H^2+\delta\mu_H^2$ with 
\begin{equation}
	\left.\delta
	\mu_H^2\right|_\mathrm{fermion}=-\frac{(Y^H)^2}{8\pi^2}\left( 
		m_\chi^2 \left(2-\frac{2m_\chi^2-m_\psi^2}{m_\chi^2-m_\psi^2}\ln\frac{m_\chi^2}{\mu^2}\right)
	+	m_\psi^2
	\left(2-\frac{2m_\psi^2-m_\chi^2}{m_\psi^2-m_\chi^2}\ln\frac{m_\psi^2}{\mu^2}\right)
	\right)
\end{equation}
Thus it receives a quadratic correction from the new fermions in the loop. This
poses a naturalness problem. The correction to the quartic coupling is of order
$(Y^H)^4$, but does not receive a quadratic enhancement by the large mass hierarchy.

Similarly the Higgs couples to the scalar $\phi$ leading to a one-loop
correction to the Higgs mass, which is shown in Fig.~\ref{fig:OneLoopPhi}.
This quartic coupling, however, is unrelated to neutrino
mass and could in principle be arbitrarily small at a given renormalization scale.
Thus the two-loop contribution is effectively the leading order contribution
related to $\phi$, that we take into account. 

In the unbroken phase its main contribution is given by diagrams of
the type shown in Fig.~\ref{fig:TwoLoopHiggsBilinear} with W bosons and $\phi$
in the loop, which we estimate as follows 
\begin{equation} 
	\left.\delta \mu_H^2\right|_\mathrm{scalar} \simeq g^4 C
	\left(\frac{1}{16\pi^2}\right)^2  A_0(m_\phi^2) =
	\frac{C\,\alpha_2^2}{16\pi^2}m_\phi^2
	\left(\frac1\epsilon+1-\ln\frac{m_\phi^2}{\mu^2}\right) \;.
\end{equation}
The constant $C$ is an order one factor, which we do not explicitly determine.
We will naively set it to $1$ in the following for simplicity, which is enough
for an order of magnitude estimate.

Evaluating the expression for the Higgs mass at the scale $\mu=m_h$
\begin{align}
m_h^2 = 4\,\mu_{H,eff}^2 =
	4\,\mu_H^2
	&-\frac{(Y^H)^2}{2\pi^2}\left( 
		m_\chi^2
		\left(2-\frac{2m_\chi^2-m_\psi^2}{m_\chi^2-m_\psi^2}\ln\frac{m_\chi^2}{m_h^2}\right)
	+	m_\psi^2
	\left(2-\frac{2m_\psi^2-m_\chi^2}{m_\psi^2-m_\chi^2}\ln\frac{m_\psi^2}{m_h^2}\right)
	\right)\nonumber\\
	&+	\frac{C\,\alpha_2^2}{4\pi^2} m_\phi^2
	\left(1-\ln\frac{m_\phi^2}{m_h^2}\right) 
\end{align}
The required fine-tuning to arrange for the correct Higgs mass by canceling the
finite correction with the term tree-level term $\mu_H^2$ can be estimated by
$\Delta^{-1}$ with  
\begin{equation}\label{eqn:ft}
	\Delta^2\equiv \sum_{i=\chi,\psi,\phi}\left(\frac{\partial\ln
	m_h^2}{\partial\ln  m_i^2} \right)^2 
	=
	\sum_{i=\chi,\psi,\phi}\left(\frac{m_i^2}{m_h^2}\frac{\partial
	m_h^2}{\partial  m_i^2} \right)^2\
	\;, 
\end{equation}
which quantifies the amount of tuning required to obtain the correct Higgs mass.

\section{Discussion}
\label{sec:other}
Currently the LUX dark matter experiment~\cite{Akerib:2013tjd} places the
strongest limit on the dark matter direct detection cross section. Despite the
mixing with $\psi$, there is no tree-level contribution to the spin-independent
cross section from $Z$-boson exchange due to the Majorana nature of the dark
matter candidate. Thus the dominant contribution arises from loop-level
processes~\cite{Cirelli:2009uv,Cai:2015zza}.
The limit on dark matter direct detection cross section is given up to 1 TeV,
which is roughly $10^{-44} \; \mathrm{cm}^2$.
If we  extrapolate the limit to the quintuplet mass, $9.4$ TeV, the limit will be much weaker 
and thus will not put any constraint on this model at the moment. This model, however, will be probed by 
future direct detection experiments such as XENON 1T~\cite{Aprile:2015uzo}.

Dark matter annihilation in our galaxy produces high energy gamma rays. 
Experiments such as High Energy Stereoscopic System (H.E.S.S)~\cite{Bernloehr:2003vd} and  the planned Cherenkov Telescope Array (CTA)~\cite{Consortium:2010bc} 
are searching for such signals and can place limits on the model.
This limit on quintuplet has been discussed in Refs.~\cite{Ostdiek:2015aga, Cirelli:2015bda, Garcia-Cely:2015dda} thoroughly. It is in tension with the current observation 
for a cuspy profile of dark matter halo, but allowed for a cored one. Nevertheless, this model is within the reach of the future CTA  independent of the dark matter profile
and will be tested soon. 

The possibility to test MDM at a collider has been extensively studied for the LHC, the HL-LHC and even a 100 TeV proton-proton collider~\cite{Cirelli:2014dsa, Low:2014cba}.    
Generally electroweak multiplets can be tested in events with mono-jet, mono-photon, vector-boson fusion and disappearing tracks. 
Among them, the test with disappearing tracks has the best sensitivity, with a reach of about 3 TeV for a 100 TeV proton-proton collider for 
an integrated luminosity of 30 $ab^{-1}$, which is still far below the mass of the dark matter candidate of this model.

\begin{figure}[btp!]\centering
		\includegraphics[width=0.8\linewidth]{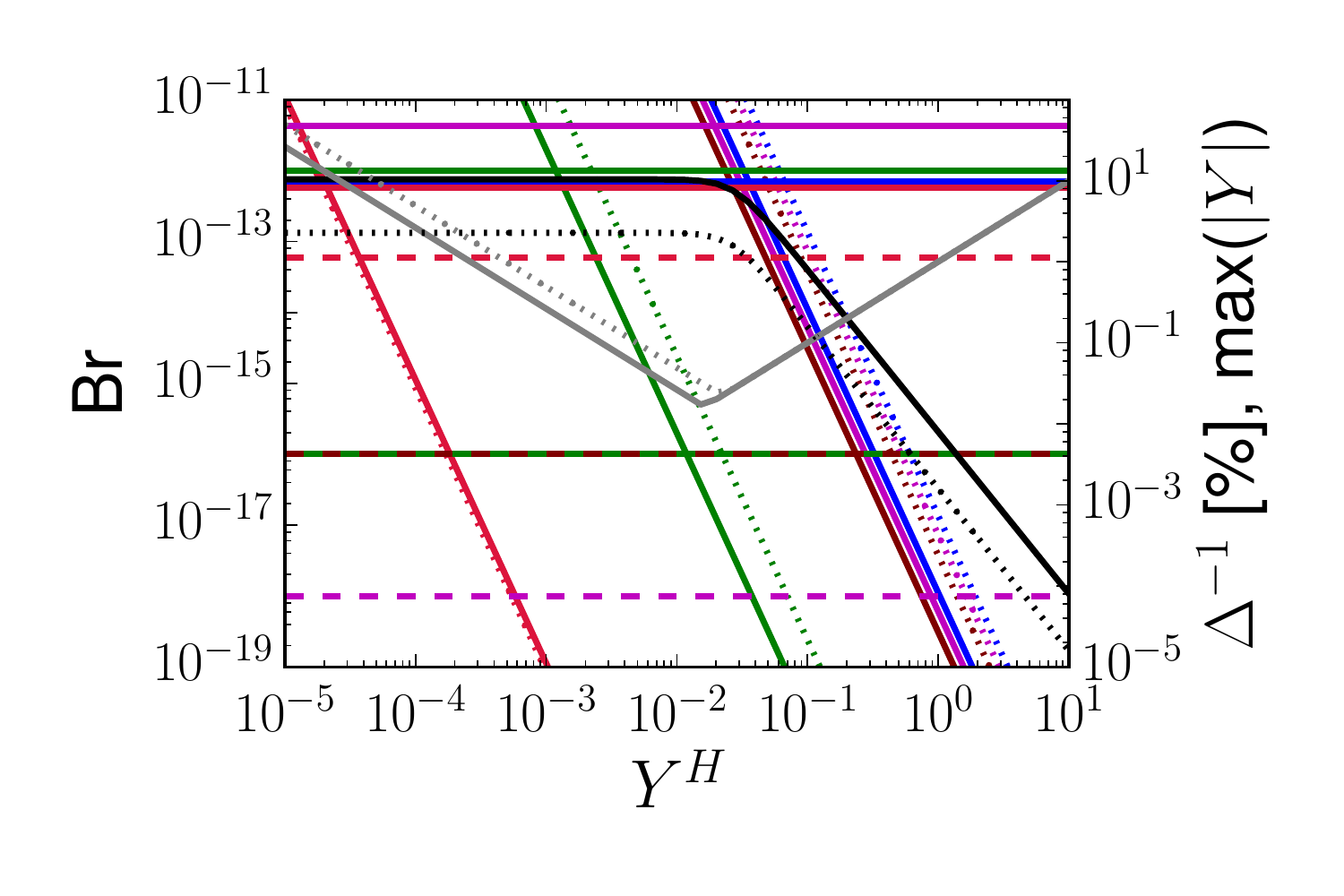}
		\caption{Prediction for lepton flavor violating processes $\mu\to e
		\gamma$ (red) , $\mu\to eee$ (green), and $\mu$-$e$ 
		conversion in gold (blue), aluminium (maroon) and titanium (magenta). The horizontal solid lines
	indicate the current experimental bound, while the dashed line indicates
	the future sensitivity of proposed experiments.
	The required fine-tuning $\Delta^{-1}$ is shown in black and the maximum
	value of the Yukawa couplings, max($Y^{L}_{ij},Y^H$), in gray.
	The solid lines are for $m_\chi=9.4$ TeV, $m_\psi=15$
TeV, $m_{\phi_1}=16$ TeV, and $m_{\phi_2}=16.5$ TeV, while the dotted lines indicate
the change if the masses of $\psi$ and $\phi_i$ are doubled. The leptonic mixing
parameters and neutrino mass squared differences are fixed to their best-fit
values. We use v2.0 of the nu-fit collaboration~\cite{Gonzalez-Garcia:2014bfa}. Leptonic CP
phases and the complex angle $\theta$ in $\mathcal{O}$ are set to zero.}\label{fig:br-yH}
	\end{figure}
	\begin{figure}[tbp!]\centering
		\includegraphics[width=0.8\linewidth]{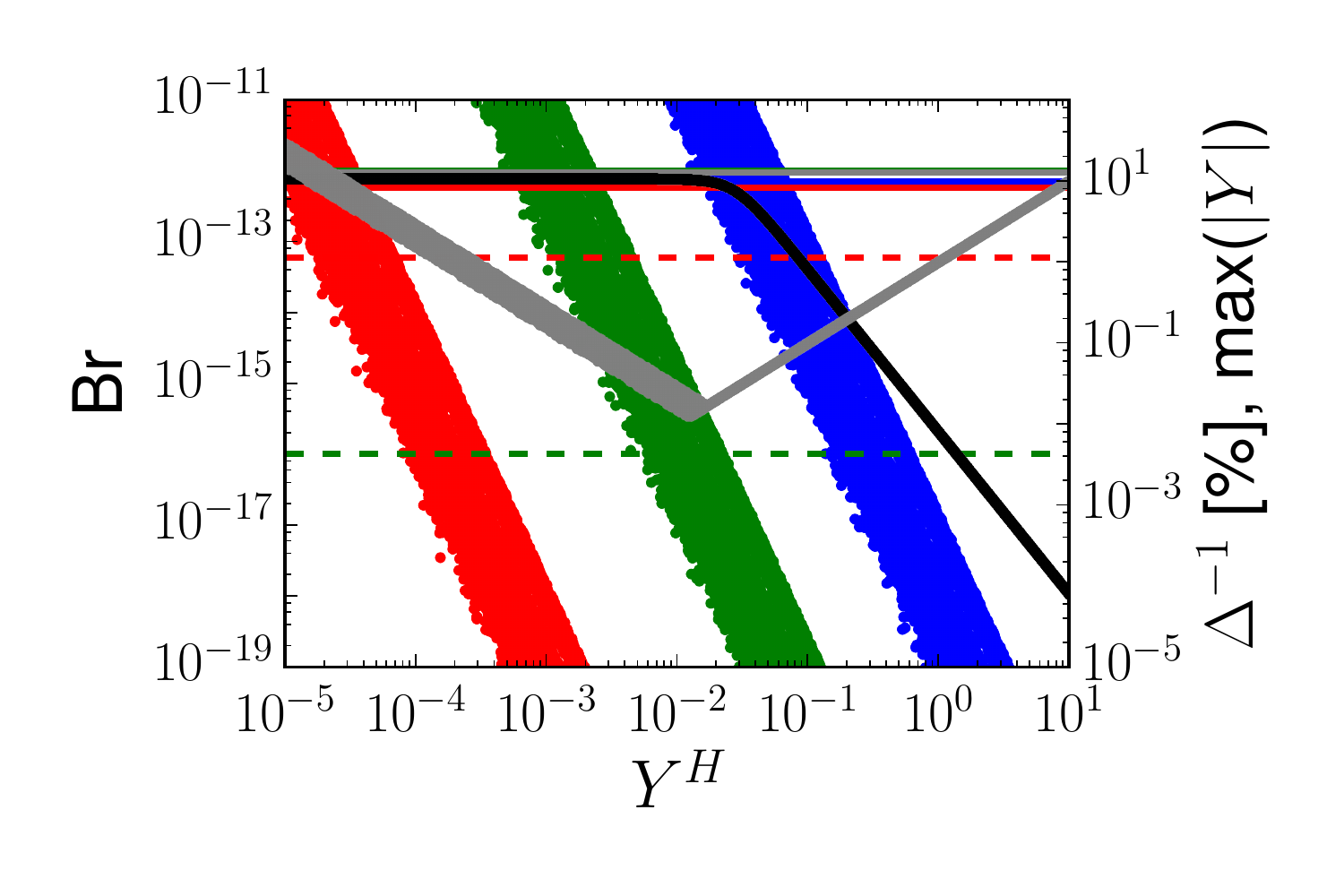}
		\caption{Same as Fig.~\ref{fig:br-yH}. However the leptonic
		mixing parameters and neutrino mass squared differences are
	varied within the allowed $3\sigma$ ranges. }\label{fig:br-yH-nu}
	\end{figure}
	\begin{figure}[tbp!]\centering
		\includegraphics[width=0.8\linewidth]{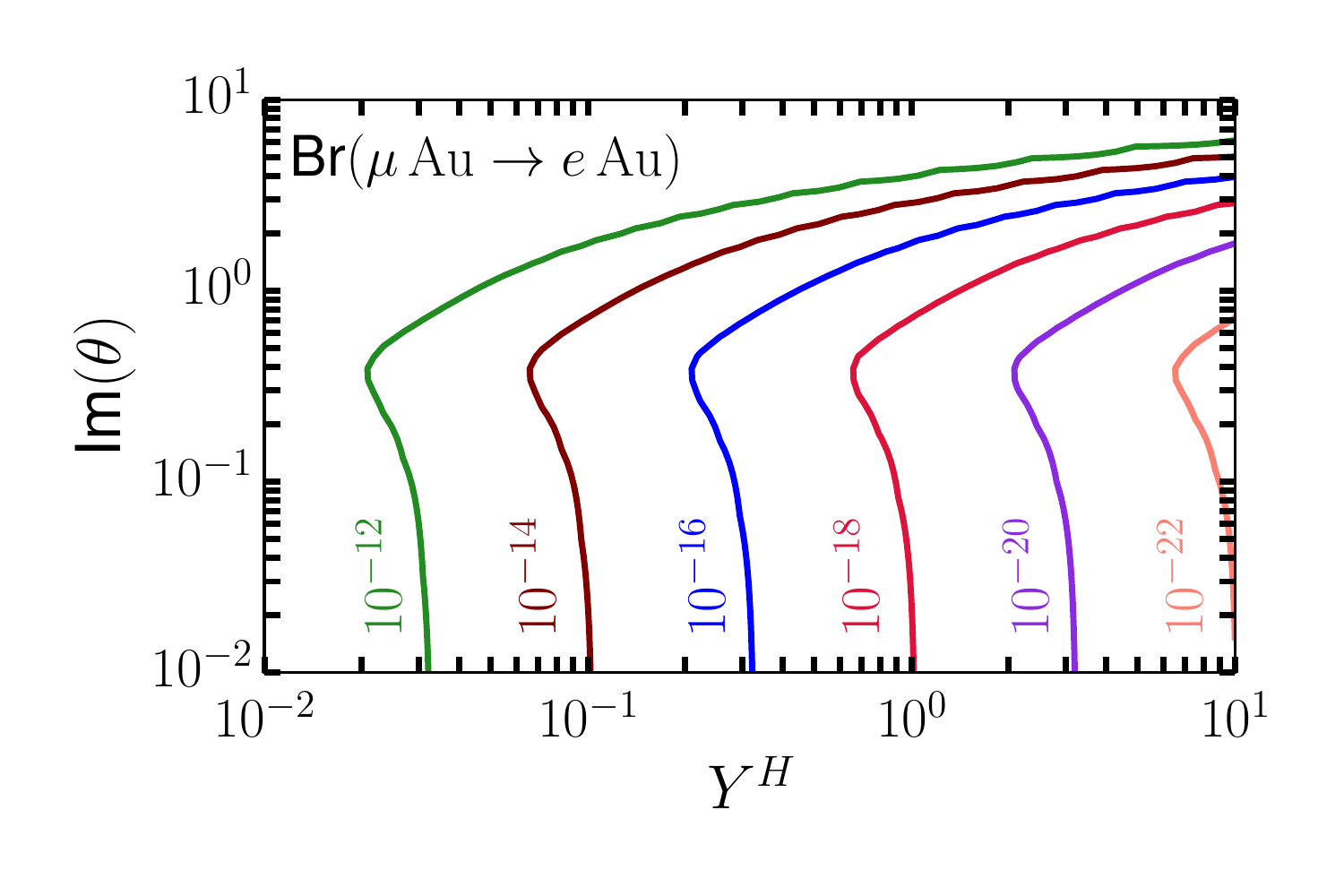}
		\caption{The contour lines show the branching ratio of $\mu$-$e$
		conversion in gold calculated using meson exchange mediation as
	a function of the Yukawa coupling $Y^H$ and the imaginary part of the
	angle $\theta$ parameterizing 
	$\mathcal{O}$.}\label{fig:br-yH-mu2e}
	\end{figure}
Lepton flavor violating processes are governed by the Yukawa coupling $Y^L$,
while the Higgs to diphoton branching ratio and naturalness are controlled by
$Y^H$. Finally, neutrino mass depends on the product of Yukawa couplings $Y^H
Y^L$ and thus connects the phenomenology creating an interesting interplay. The
smaller $Y^H$, the larger $Y^L$ and consequently the larger the LFV branching
ratios. In Fig.~\ref{fig:br-yH} we show the branching ratio for the lepton flavor violating processes $\mu\to e
		\gamma$ (red) , $\mu\to eee$ (green), and $\mu$-$e$ 
		conversion\footnote{We use the values of Tab.~1 of
		Ref.~\cite{Kitano:2002mt} for the overlap integrals. A
		comparison of the rates for the overlap integral values in
		Tabs.~2 and 4 of \cite{Kitano:2002mt} indicates an uncertainty of about
		44\%, 5\%, and 11\% for Au, Al, and Ti, respectively.} in gold (blue), aluminium (maroon), and titanium (magenta) as a function
		of $Y^H$ for fixed masses $m_\chi=9.4$ TeV, $m_\psi=15$
TeV, $m_{\phi_1}=16$ TeV, and $m_{\phi_2}=16.5$ TeV as solid lines. The dotted
lines indicate the change if the masses of $\psi$ and $\phi_i$ are doubled.
The solar and atmospheric mass squared differences and the
		leptonic mixing parameters are fixed to the best-fit values of
		v2.0 of the nu-fit collaboration~\cite{Gonzalez-Garcia:2014bfa}. Leptonic CP
		phases are set to zero, as well as the complex angle $\theta$
		parameterizing $\mathcal{O}$.
		The horizontal solid lines show the current experimental
		bounds on Br($\mu\to e\gamma)<5.7\times
		10^{-13}$~\cite{Adam:2013mnn} (red), Br($\mu\to
		eee)<10^{-12}$~\cite{Bellgardt:1987du} (green), 
		Br$(\mu\mathrm{Ti}\to e
		\mathrm{Ti})<4\times
		10^{-13}$~\cite{Dohmen:1993mp} (magenta),
		and
		Br$(\mu\mathrm{Au}\to e
		\mathrm{Au})<7\times
		10^{-13}$~\cite{Bertl:2006up} (blue), while the dashed lines indicate
		the future sensitivities: Br($\mu\to
		e\gamma)\sim 6\times 10^{-14}$~\cite{Baldini:2013ke}, Br($\mu\to eee)\sim
		10^{-16}$ ~\cite{Blondel:2013ia}, Br($\mu \mathrm{Al} \to e
		\mathrm{Al})\sim 10^{-16}$
	\cite{Hungerford:2009zz,Cui:2009zz,Carey:2008zz,Kurup:2011zza,Kutschke:2011ux})
	and  Br($\mu \mathrm{Ti} \to e \mathrm{Ti})\sim 10^{-18}$~\cite{Hungerford:2009zz, Cui:2009zz}
		For fixed masses and Yukawa coupling $Y^H$, it is possible to
		quantify how much the parameters of the model have to be tuned
		to obtain the electroweak scale. We show the required fine-tuning
		$\Delta^{-1}$, which is defined in Eqn.~\eqref{eqn:ft}, as a black solid line for our benchmark point and dotted
	line for doubled particle masses. The values can be read off the
	y-axis on the right-hand side. The Yukawa coupling $Y^H$ becomes
	non-perturbative on the right-hand side of the figure, while the largest
	entry of the Yukawa coupling $Y^L$ becomes non-perturbative on the
	left-hand side of the figure. We plot the maximum value of the Yukawa
couplings, max$(Y^{L}_{ij},Y^H)$, in gray. The value can be read off the y-axis on the right-hand side.

	Fig.~\ref{fig:br-yH-nu} illustrates the uncertainty in the solar and
	atmospheric mass squared differences and leptonic mixing parameters. All
	these parameters are varied within their
	$3\sigma$ allowed ranges. The different colors and line
	styles are chosen in the same way as in Fig.~\ref{fig:br-yH}. Fixing
	everything but the leptonic mixing parameters and mass squared
	differences, there is an
	uncertainty of up to three orders of magnitude in the rates for the
	different processes.
	Fig.~\ref{fig:br-yH-mu2e} shows a contour plot of Br($\mu
	\mathrm{Au} \to e \mathrm{Au}$) in the plane of the imaginary part of
	the complex angle $\theta$ in the complex orthogonal matrix
	$\mathcal{O}$ and the Yukawa coupling $Y^H$. All LFV rates are
	relatively insensitive to the real part of $\theta$ and change at most at the
	percent level. A large imaginary part however leads to large Yukawa 
	couplings $Y^L$ canceling among each other to accommodate the
	light neutrino mass.
	
Finally we want to comment on the renormalization group evolution of the
couplings in this model. The large SU(2) representations lead to a strong running
of the gauge couplings resulting in a Landau pole at a scale of about $10^9$
GeV, where we have taken two-loop running into account. Similarly the quartic
couplings of large scalar SU(2) representations suffer from the triviality
bound~\cite{Cabibbo:1979ay,Lindner:1986uk,Hamada:2015bra}, because gauge
couplings will induce these couplings at one-loop order, which will be further amplified by the
running of the respective quartic coupling itself~\cite{Hamada:2015bra}.  In particular,
the study in Ref.~\cite{Hamada:2015bra} finds that the quartic coupling of the
real quintuplet suffers from a Landau pole below
$10^{15}$ GeV. A viable UV completion has to preserve the accidental $Z_2$
symmetry to prevent the minimal dark matter candidate from
decaying. We are agnostic about the UV completion and mainly concentrate on
phenomenology, since there are good prospects to ultimately test minimal dark
matter models in the near future.

\section{Conclusion}
\label{sec:con}
Embedding radiative neutrino mass models in the framework of MDM to make one theory work for two major fields of physics beyond the SM 
is aesthetically appealing. We systematically studied the possibility to realize this idea with
radiative neutrino mass models as UV completions of the Weinberg operator. None
of the minimal UV completions at one-loop leads to a stable minimal dark matter candidate.
However we argued that it is feasible to obtain a cosmologically stable dark
matter candidate, because the decay is controlled by a coupling unrelated to
neutrino mass generation, which can naturally be arbitrarily small.

We studied the phenomenology of one model explicitly. The model contains two
real quintuplet scalars and also a quintuplet fermion whose neutral component field plays the role of dark matter. 
Both fields have zero hypercharge. In addition, a vector-like quadruplet fermion are introduced with hypercharge $\pm \frac{1}{2}$.
We discussed the neutrino mass generation in this model and performed a detailed
phenomenological study of lepton flavour violation and Higgs decay. There is a sizable allowed region of parameter space
consistent with all current experimental constraints from Higgs physics and lepton flavor changing processes.  
The most stringent bound is placed by $\mu$-$e$ conversion in nuclei and will be
further improved by future experiments. It places a lower bound on the Yukawa
coupling $Y^H$ and thus increases the electroweak fine-tuning. Current
bounds already require at least $\sim10\%$ tuning. In the near future, the
remaining parameter space of this
model can be tested by direct detection experiments like XENON 1T and indirect
detection experiments such as CTA. 

\section*{Acknowledgments}
We thank Diego Aristizabal Sierra for pointing out the loop-induced decay modes of the
minimal dark matter candidates via private communication and for useful discussions.
 We acknowledge the use of \texttt{matplotlib}~\cite{Hunter:2007} and \texttt{ipython}~\cite{PER-GRA:2007}.
 This work was supported in part by the Australian Research Council.  
 
 \appendix

\section{Mass Matrix Diagonalization}
\label{app:diagonalization}
The mass mixing in this model is proportional to  $Y^H v/m_{\psi,\chi}$ and thus very suppressed for heavy masses of $\mathcal{O}(10)$ TeV. 
Although it can be safely neglected in most of the calculation as we did, we show the technical details for the diagonalization of the mass matrices for completeness.

The mass matrix for the neutral fermions can be diagonalised using a Takagi factorization, $V_0^* M_0 V_0^\dagger = M_0^D$. At leading order we find   
\begin{align}
	M_0^D&= \begin{pmatrix}
	m_\psi &&\\
	       &m_\psi&\\
	       &&m_\chi\\
\end{pmatrix}\;,&
	V_0&=
	\left(
\begin{array}{ccc}
 -\frac{i}{\sqrt{2}} & \frac{i}{\sqrt{2}} & \frac{i v Y^H}{2 \sqrt{2} \left(m_{\chi
   }+m_{\psi }\right)} \\
 \frac{1}{\sqrt{2}} & \frac{1}{\sqrt{2}} & -\frac{v Y^H}{\sqrt{2} \left(2 m_{\chi }-2
   m_{\psi }\right)} \\
 \frac{v Y^H m_{\chi }}{2 m_{\chi }^2-2 m_{\psi }^2} & \frac{v Y^H m_{\psi }}{2 m_{\chi
   }^2-2 m_{\psi }^2} & 1 \\
\end{array}
\right)\;.
\end{align}
For the singly-charged fermions, the mass matrix $M_1$ can be diagonalised with a singular value decomposition
$V_1^* X_1 W_1^\dagger = X_1^D$ with a diagonal matrix $X_1^D$ and two unitary
matrices $V_1$, $W_1$. To the leading order they are given by
\begin{align}
	X_1^D&=\begin{pmatrix}
	m_\psi && \\
	       &m_\psi & \\
		      && m_\chi \\
\end{pmatrix}\;,
&
	V_1&=
\left(
\begin{array}{ccc}
 \frac{1}{\sqrt{2}} & -\frac{1}{\sqrt{2}} & \frac{v Y^H \left(\sqrt{3} m_{\chi }+m_{\psi
   }\right)}{4 \left(m_{\chi }^2-m_{\psi }^2\right)} \\
 -\frac{i}{\sqrt{2}} & -\frac{i}{\sqrt{2}} & -\frac{i v Y^H \left(\sqrt{3} m_{\chi
   }-m_{\psi }\right)}{4 \left(m_{\chi }^2-m_{\psi }^2\right)} \\
 -\frac{\sqrt{\frac{3}{2}} v Y^H m_{\chi }}{2 m_{\chi }^2-2 m_{\psi }^2} & \frac{v Y^H
   m_{\psi }}{\sqrt{2} \left(2 m_{\chi }^2-2 m_{\psi }^2\right)} & 1 \\
\end{array}
\right)\;,
\end{align}
\begin{align}
W_1&=
\left(
\begin{array}{ccc}
 \frac{1}{\sqrt{2}} & -\frac{1}{\sqrt{2}} & \frac{v Y^H \left(m_{\chi }+\sqrt{3} m_{\psi
   }\right)}{4 \left(m_{\chi }^2-m_{\psi }^2\right)} \\
 -\frac{i}{\sqrt{2}} & -\frac{i}{\sqrt{2}} & -\frac{i v Y^H \left(m_{\chi }-\sqrt{3}
   m_{\psi }\right)}{4 \left(m_{\chi }^2-m_{\psi }^2\right)} \\
 -\frac{v Y^H m_{\chi }}{\sqrt{2} \left(2 m_{\chi }^2-2 m_{\psi }^2\right)} &
   \frac{\sqrt{\frac{3}{2}} v Y^H m_{\psi }}{2 m_{\chi }^2-2 m_{\psi }^2} & 1 \\
\end{array}
\right)
\; .
\end{align}
Similarly for the doubly-charged fermions, the mass matrix $M_2$ can be diagonalised with a singular value decomposition
$V_2^* X_2 W_2^\dagger = X_2^D$ with a diagonal matrix $X_2^D$ and two unitary
matrices $V_2$, $W_2$. To leading order they are given by
\begin{align}
	X_2^D&=\begin{pmatrix}
		m_\psi & \\
		       & m_\chi \\
\end{pmatrix}\;,\\
	V_2&=\left(
\begin{array}{cc}
 1 & -\frac{v Y^H m_{\chi }}{\sqrt{2} \left(m_{\chi }^2-m_{\psi }^2\right)} \\
 \frac{v Y^H m_{\chi }}{\sqrt{2} \left(m_{\chi }^2-m_{\psi }^2\right)} & 1 \\
\end{array}
\right)\;,&
W_2&=\left(
\begin{array}{cc}
 1 & \frac{v Y^H m_{\psi }}{\sqrt{2} \left(m_{\psi }^2-m_{\chi }^2\right)} \\
 \frac{v Y^H m_{\psi }}{\sqrt{2} \left(m_{\chi }^2-m_{\psi }^2\right)} & 1 \\
\end{array}
\right)\;.
\end{align}

 \section{SU(2)$_L$ generators and the kinetic terms}
\label{app:su2}
All the odd-dimensional representations are real and even-dimensional
representations pseudo-real. The generators of the four-dimensional
representations can be explicitly written as
\begin{align}
J_1^4  &=\left(
 \begin{tabular}{cccc}
  0 & $-\frac{\sqrt{3}}{2}$ & 0 & 0\\
  $-\frac{\sqrt{3}}{2}$ & 0 & $-1$ & 0\\
  0  & $-1$ & 0 & $\frac{\sqrt{3}}{2}$ \\
  0 & 0 & $\frac{\sqrt{3}}{2}$ & 0 
 \end{tabular}
 \right)\;, 
& J_2^4 & = i \left(
 \begin{tabular}{cccc}
  0 & $\frac{\sqrt{3}}{2}$ & 0 & 0\\
  $-\frac{\sqrt{3}}{2}$ & 0 & 1 & 0\\
  0  & $-1$ & 0 & $-\frac{\sqrt{3}}{2}$ \\
  0 & 0 & $\frac{\sqrt{3}}{2}$ & 0 
 \end{tabular} 
\right)\;, \\\nonumber
	J_3^4 & = \rm{diag} (\frac{3}{2}, \frac{1}{2}, -\frac{1}{2}, -\frac{3}{2}) 
\end{align}
and the generators of the five-dimensional representation are given by
\begin{align}
J_1^5&=\left(
\begin{tabular}{ccccc}
0 & $-1$ & 0 & 0 & 0 \\
$-1$ & 0 & $-\sqrt{\frac{3}{2}}$ & 0 & 0 \\
0 & $-\sqrt{\frac{3}{2}}$ & 0 & $\sqrt{\frac{3}{2}}$ & 0 \\
0 & 0 & $\sqrt{\frac{3}{2}}$ & 0 & 1 \\
0 & 0 & 0 & 1 & 0 
\end{tabular}
\right)\;,
&
J_2^5 &= i \left(
\begin{tabular}{ccccc}
0 & $1$ & 0 & 0 & 0 \\
$-1$ & 0 & $\sqrt{\frac{3}{2}}$ & 0 & 0 \\
0 & $-\sqrt{\frac{3}{2}}$ & 0 & $-\sqrt{\frac{3}{2}}$ & 0 \\
0 & 0 & $\sqrt{\frac{3}{2}}$ & 0 & $-1$ \\
0 & 0 & 0 & 1 & 0 
\end{tabular}
\right)\;,\\\nonumber
J_3^5 & = \rm{diag}\left( 2, 1, 0, -1, -2\right)
 \; .
\end{align}
The kinetic terms for the exotic field are expressed as
\begin{align}
\mathcal{L}^{kin}&=\frac{1}{2}\left( D_\mu \phi\right)^\dagger D^\mu \phi 
          + i \chi^\dagger \bar{\sigma}^\mu D_\mu \chi 
       + i \psi^\dagger \bar{\sigma}^\mu D_\mu \psi  + i \bar{\psi}^\dagger
       \bar{\sigma}^\mu D_\mu \bar{\psi} \\\nonumber
     &  \quad     - \frac{1}{2}m_\phi^2 \phi^\dagger \phi 
            -\frac{1}{2} m_\chi (\chi \chi +\chi^\dagger\chi^\dagger)  -m_\psi
	    (\psi \bar{\psi} + \psi^\dagger \bar{\psi}^\dagger)\\\nonumber
& = \left( D_\mu \phi\right)^\dagger D^\mu \phi + i \chi^\dagger_i \bar{\sigma}^\mu D_\mu \chi_i  
       + i \psi^\dagger \bar{\sigma}^\mu D_\mu \psi  + i \bar{\psi}^\dagger
       \bar{\sigma}^\mu D_\mu \bar{\psi} \\\nonumber
& \quad  - m_\phi^2 \left(\frac{1}{2}\phi_0^2 + \phi^+\phi^-
+\phi^{++}\phi^{--}\right)  \\\nonumber
& \quad  - m_\chi \left(\frac{1}{2}\chi^0\chi^0  - \chi^-\chi^+ +
\chi^{--}\chi^{++} + h.c.\right)  \\\nonumber
& \quad  - m_\psi \left( \psi^0\bar{\psi}^0 - \psi^+\bar{\psi}^- - \psi^-\bar{\psi}^+ + \psi^{++}\bar{\psi}^{--} + h.c. \right)\; ,
\end{align}
where the covariant derivatives are $D_\mu = \partial_\mu - i g J_a W_\mu^a - i g^\prime Y B_\mu$ .  

\bibliography{ref}{}

\end{document}